\documentclass[twocolumn]{aastex62}

\graphicspath{{./}{figures/}}

\received{January 1, 2018}
\revised{January 7, 2018}
\accepted{\today}
\submitjournal{ApJ}

\shorttitle{Dust Lanes In Edge-On Galaxies in GAMA/KiDS}
\shortauthors{Holwerda et al.}

\begin{document}

\title{The frequency of dust lanes in edge-on spiral galaxies  identified by Galaxy Zoo in KiDS imaging of GAMA targets}

\correspondingauthor{Benne W. Holwerda}
\email{benne.holwerda@loisville.edu}

\author[0000-0002-4884-6756]{Benne W. Holwerda}
\affil{Department of Physics and Astronomy, 102 Natural Science Building, University of Louisville, Louisville KY 40292, USA}

\author{Lee Kelvin}
\affil{Astrophysics Research Institute, Liverpool John Moores University, IC2, Liverpool Science Park, 146 Brownlow Hill, Liverpool, L3 5RF, United Kingdom}

\author{Ivan Baldry}
\affil{Astrophysics Research Institute, Liverpool John Moores University, IC2, Liverpool Science Park, 146 Brownlow Hill, Liverpool, L3 5RF, United Kingdom}

\author{Chris Lintott}
\affil{Denys Wilkinson Building
Keble Road
Oxford
OX1 3RH, United Kingdom}

\author{Mehmet Alpaslan}
\affil{Center for Cosmology and Particle Physics
Department of Physics, New York University
726 Broadway, Office 913
New York, NY 10003, USA}

\author{Kevin A Pimbblet}
\affil{E.A.Milne Centre for Astrophysics, University of Hull, Cottingham Road, Kingston-upon-Hull, HU6 7RX, UK}

\author{Jochen Liske}
\affil{Universit\"at Hamburg, 
Hamburger Sternwarte, 
Gojenbergsweg 112, 
21029 Hamburg, Germany}

\author{Thomas Kitching}
\affil{Mullard Space Science Laboratory, University College London, Holmbury St. Mary, Dorking, Surrey RH5 6NT, UK}

\author{Steven Bamford}
\affil{University of Nottingham, University Park, Nottingham, NG7 2RD, United Kingdom} 

\author{Jelte de Jong}
\affil{Leiden Observatory, Universiteit Leiden, Niels Bohrweg 2,NL-2333 CA Leiden, The Netherlands}

\author{Maciej Bilicki}
\affil{Leiden Observatory, Universiteit Leiden, Niels Bohrweg 2,NL-2333 CA Leiden, The Netherlands} 

\author{Andrew Hopkins}
\affil{Australian Astronomical Observatory
105 Delhi Rd, North Ryde, NSW 2113, Australia}

\author{Joanna Bridge}
\affil{Department of Physics and Astronomy, 102 Natural Science Building, University of Louisville, Louisville KY 40292, USA} 

\author{R. Steele}
\affil{Department of Physics and Astronomy, 102 Natural Science Building, University of Louisville, Louisville KY 40292, USA} 

\author{A. Jacques}
\affil{Department of Physics and Astronomy, 102 Natural Science Building, University of Louisville, Louisville KY 40292, USA} 

\author{S. Goswami}
\affil{Department of Physics and Astronomy, 102 Natural Science Building, University of Louisville, Louisville KY 40292, USA} 

\author{S. Kusmic}
\affil{Department of Physics and Astronomy, 102 Natural Science Building, University of Louisville, Louisville KY 40292, USA} 

\author{W. Roemer}
\affil{Department of Physics and Astronomy, 102 Natural Science Building, University of Louisville, Louisville KY 40292, USA} 

\author{S. Kruk}
\affil{University of Oxford, Denys Wilkinson Building, Keble Road, Oxford, OX1 3RH, United Kingdom} 

\author{C.C. Popescu}
\affil{Jeremiah Horrocks Institute, University of Central Lancashire, Preston, United Kingdom} 
\affil{The Astronomical Institute of the Romanian Academy, Str. Cutitul de Argint 5, Bucharest, Romania}

\author{K. Kuijken}
\affil{Leiden Observatory, Universiteit Leiden, Niels Bohrweg 2,NL-2333 CA Leiden, The Netherlands} 

\author{L. Wang}
\affil{SRON Netherlands Institute for Space Research, Landleven 12, 9747 AD, Groningen, The Netherlands}
\affil{Kapteyn Astronomical Institute, University of Groningen, Postbus 800, 9700 AV, Groningen, The Netherlands}

\author{A. Wright}
\affil{Argelander-Institut f\"ur Astronomie (AIfA) Universit\"at Bonn
Auf dem H\"ugel 71 D-53121 Bonn, Germany}

\begin{abstract}

Dust lanes bisect the plane of a typical edge-on spiral galaxy as a dark optical absorption feature. Their appearance is linked to the gravitational stability of spiral disks; the fraction of edge-on galaxies that displays a dust lane is a direct indicator of the typical vertical balance between gravity and turbulence; a balance struck between the energy input from star-formation and the gravitational pull into the plane of the disk.

Based on morphological classifications by the Galaxy~Zoo project on the Kilo-Degree Survey (KiDS) imaging data in the Galaxy and Mass Assembly (GAMA) fields, we explore the relation of dust lanes to the galaxy characteristics, most of which were determined using the {\sc magphys} spectral energy distribution fitting tool: stellar mass, total and specific star-formation rates, and several parameters describing the cold dust component.

We find that the fraction of dust lanes does depend on the stellar mass of the galaxy; they start to appear at $M^* \sim 10^9 M_\odot$. A dust lane also implies strongly a dust mass of at least $10^5 M_\odot$, but otherwise does not correlate with cold dust mass parameters of the {\sc magphys} spectral energy distribution analysis, nor is there a link with star-formation rate, specific or total.  Dust lane identification does not depend on disk ellipticity (disk thickness) or Sersic profile but correlates with bulge morphology; a round bulge favors dust lane votes.

The central component along the line of sight that produces the dust lane is not associated with either one of the components fit by {\sc magphys}, the cold diffuse component or the localized, heated component in HII regions, but a mix of these two.

\end{abstract}



\section{Introduction} \label{sec:intro}

A dark stripe in the mid-plane of the spiral disk is part of the canonical edge-on view of late-type galaxies. These dust lanes are so common that their presence is often taken as a signature of a perfectly edge-on disk (inclination $i>85^{\circ}$).
\cite{Dalcanton04} show that dust lanes appear predominantly in massive galaxies ($v_{\rm rot}>120$~km/s or a stellar mass of $\sim 10^{9.8} M_\odot$). They link the phenomenon to the vertical stability -- the Toomre $Q$ criterion \citep{Toomre64} -- of the gas and stellar spiral disk that hosts the dust lanes: if the surface density is sufficiently high, the disk vertically collapses into a thin disk. In smaller galaxies, the interstellar matter (ISM) is relatively more distributed throughout the height of the stellar disk, i.e., the amount of dusty ISM is the same relative to the stellar mass but is not concentrated in the plane to form the line-of-sight dust lane seen in the edge-on disk.
However, the \cite{Dalcanton04} sample is small (49~galaxies) and is made up of predominantly bulge-less galaxies. \cite{Obric06} did an initial pass on the SDSS galaxies and found that the fraction of dust lanes dramatically increased at $v_{\rm rot} = 150$~km/s for all late-types.
Both these studies point to a fundamental change in spiral disks with halo or stellar mass. At a critical halo size, the disk flattens conspicuously with respect to it's size -- the ISM more so than the stellar disk. This has implications for the observed global galaxy characteristics: a condensed ISM disk may form stars more efficiently, the vertical instability affects the spiral density wave, the formation efficiency of bars may change, a compact dusty ISM lowers the UV photon escape fraction  \citep[e.g.][]{Dijkstra12,Stark15,Dijkstra14,Dijkstra16,Bridge18}. If the transition is indeed sudden, it constitutes a fundamental phase change in the ISM of spirals. 

In emission, the picture should be clearer: there is no need for a stellar disk to backlight the 
dust structures. Thus, sub-mm observations with {\em Herschel} and {\em Spitzer} of vertically resolved, edge-on disks should reveal if 
there does exist a sharp transition in the dusty ISM structure. Several programs with {\em Herschel} target massive edge-on spirals, notably the HEROES project \citep{Verstappen13}. The NHEMESES program 
\citep[][Holwerda et al. {\em in preparation}]{Holwerda11iau, Holwerda12a} is designed specifically to target smaller disks to explore the dust morphology and any transition in structure. However, the height of the disk is only just resolved at the longer wavelengths, and the emission depends on the temperature of the cold dust grains. It has proven difficult to disentangle the vertical ISM density profile from the vertical temperature gradient.

The thickness of the dusty ISM disks was shown first by the Radiative Transfer (RT) models of several disks by \cite{Xilouris99}. \cite{Alton98,Alton00b} follow the initial results with more NGC 891 observations that show a large fraction of the disk has dust emission associated with it.

Building on these initial result, several groups have constructed RT models to explain multi-wavelength data of edge-on galaxies.
There are several groups using RT models (mostly of NGC 891) to map typical dust in spiral galaxies.
\cite{Popescu00,Popescu11} model NGC 891 in detail with a diffuse disk and stellar nursery components to explain its multi-wavelength behavior. \cite{Misiriotis01} follows this up with 4 more galaxies. This NGC 891 model is the basis for a correction of disk galaxy photometry etc \citep{Pastrav12,Wijesinghe11a,Wijesinghe11b,Grootes13a}.
\cite{Bianchi08} present the TRADING RT code and model NGC 891 with it \citep{Bianchi11} as well as follow-up on the initial Xilouris work \citep{Bianchi07}.
\cite{Schechtman-Rook12a,Schechtman-Rook13b} analyze NGC 891 in a completely new way, using large-scale SED models together with small-scale dust structures observed in HST fields to model the dust disk of this canonical edge-on galaxy. 

The current state-of-the-art in edge-on galaxy radiative transfer is the SKIRT fitting code \cite{Baes01a,Baes01b,Baes11}. The power of this modeling code is that is has a fitting component \citep{De-Geyter13}, making it by far the most suitable for future work and flexible enough to quickly incorporate new data.
SKIRT has resulted in models that have come closest to explaining the full multi-wavelength data on Milky-Way type edge-on galaxies \citep[HEROES Herschel project][]{de-Looze12b,Verstappen13,Allaert15,De-Geyter14,De-Geyter15,Mosenkov16}. Once again NGC 891 makes an appearance in the first SKIRT efforts \citep{Hughes14,Hughes15}. The main issue for SKIRT remains an under-prediction of the face-on optical depth \citep[e.g.][]{Holwerda05} or an under-prediction of the sub-mm emission \citep{Saftly15}

The main --generalized-- results coming out of these SKIRT efforts for massive spirals is that the dust is in a disk with a scale-height 50\% of the stellar disk's scale-height and the dust disk's scale-length is 150\% the scale-length of the stellar disk respectively, initially already reported in \cite{Xilouris99}. In addition to this diffuse disk, clumpy structures are around star-formation regions. 
This model of diffuse+clumps is used in the {\sc magphys} SED code \citep{magphys} as well. The balance between the small and large dust structure models is key to furthering our understanding of their stellar light from this point \citep{Saftly15}. A big strength is that optical measurements can be compared directly to dust emission \citep[e.g.,][]{Hughes15}.

A smaller effort is under way to characterize the disk galaxies much less massive than the Milky Way \citep[NHEMESES][\& {\em in prep}]{Holwerda12}. A remaining issue with SKIRT and all the other RT models is that the face-on central optical depth remains low in comparison with transmission measurements by a factor $\sim2$ \citep{Holwerda05} and \citep{Keel13}. 

A complementary effort is therefore to leverage the statistics of optical imaging surveys on dust lane frequency. 
We use data from the citizen science project Galaxy~Zoo\footnote{\url{http://www.Galaxy Zoo.org}} \citep{galaxyzoo} for better statistics of absorption features in late-type edge-on galaxies.

The Galaxy~Zoo project has already proven itself unparalleled in the large-scale analysis of morphological phenomena, 
until recently the purview of specialist classifiers. For example, Galaxy Zoo classifications \citep{Lintott08, Lintott11a, Fortson12} have been used to identify mergers \citep{Darg10a, Darg10b, Darg11,Casteels13}, the prevalence of bars \citep{Hoyle11, Masters11, Masters12, Kruk18}, and occulting galaxy pairs \citep{Keel13}. The identification of dusty structures in SDSS images has already proven very scientifically worthwhile: \cite{Kaviraj11} and \cite{Shabala12} show how dust structures prevail in massive elliptical galaxies. 
Here we focus on those galaxies identified by the Galaxy Zoo as disk-dominated, spiral galaxies, seen edge-on.


\cite{Holwerda12b} show that the dust lane fraction in massive $L^\star_V$ galaxies barely changes with redshift out to $z\sim0.8$. This was calibrated with a select sample of $L^\star_V$ SDSS galaxies, for which they also found a dust lane fraction of $\sim80\%$. This indicates that the dust lane is a very constant phenomenon in massive disks, if not necessarily a non-transient one -- i.e., dust lanes may still be rapidly both destroyed and re-formed. However, the edge-on view, which is often optically thick, is the most robust to such changes.
In this paper, we explore the links in the local Universe between dust lane occurrence and disk properties.

Our goals for this paper are to explore 
(a) if the sharp transition in dust lane frequency is still seen at the stellar mass that \cite{Dalcanton04} observed in bulgeless galaxies, 
(b) what the effect of a bulge is on dust lane frequency, and
(c) the relation between galaxy properties and dust lane frequency.
This paper is organized as follows: 
\S \ref{s:data} describes the sample selection from the Galaxy~Zoo database, 
\S \ref{s:gz} describes the part of the GAMA-KiDS Galaxy Zoo decision tree relevant to this project,
\S \ref{s:results} presents the results for the numbers of galaxies identified as disk, edge-on and with a dust lane as a function of various galaxy properties,
\S \ref{s:disk} briefly discusses these results and \S \ref{s:concl} lists our conclusions. 


\begin{figure}
\includegraphics[width=0.5\textwidth]{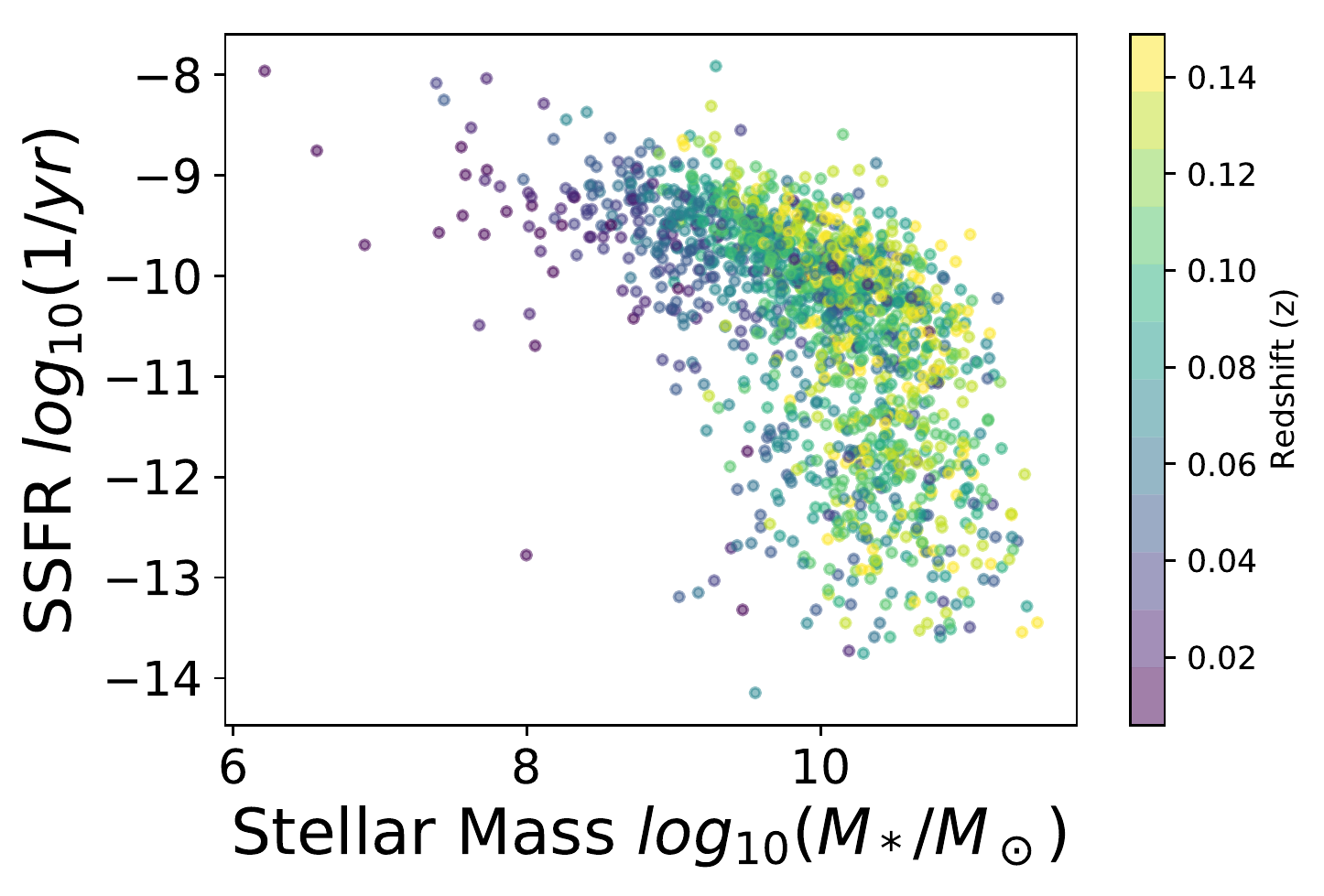}
\caption{\label{f:sm_ssfr:z} The stellar mass and specific star-formation plot of the GAMA galaxies classified by GalaxyZoo with the GAMA spectroscopic redshift for the colorbar. The detection and inclusion of low mass galaxies are biased towards high specific star-formation and low redshift. }
\end{figure}

\section{Sample Selection and Data} \label{s:data}



 The Galaxy Zoo classifications are based on the Galaxy and Mass Assembly survey DR2 and the Kilo Degree Survey (KiDS) imaging. 
For the Galaxy Zoo classification, 49851 galaxies were selected from the equatorial fields with redshifts $z < 0.15$. The Galaxy Zoo provided a monumental effort with almost 2 million classifications received from over 20,000 unique users over the course of 12 months. 

The Kilo Degree Survey \citep[KiDS][]{de-Jong13,de-Jong15,de-Jong17} is an ongoing optical wide-field imaging survey with the OmegaCAM camera at the VLT Survey Telescope. It aims to image 1350 deg$^2$ in four filters (u g r i). The core science driver is mapping the large-scale matter distribution in the Universe, using weak lensing shear and photometric redshift measurements. Further science cases include galaxy evolution, Milky Way structure, detection of high-redshift clusters, and finding rare sources such as strong lenses and quasars.
KiDS image quality is typically 0\farcs6 resolution (for sdss-r) and depths of 23.5, 25, 25.2, 24.2 magnitude for i, r, g and u respectively. 

\begin{figure}
\includegraphics[width=0.5\textwidth]{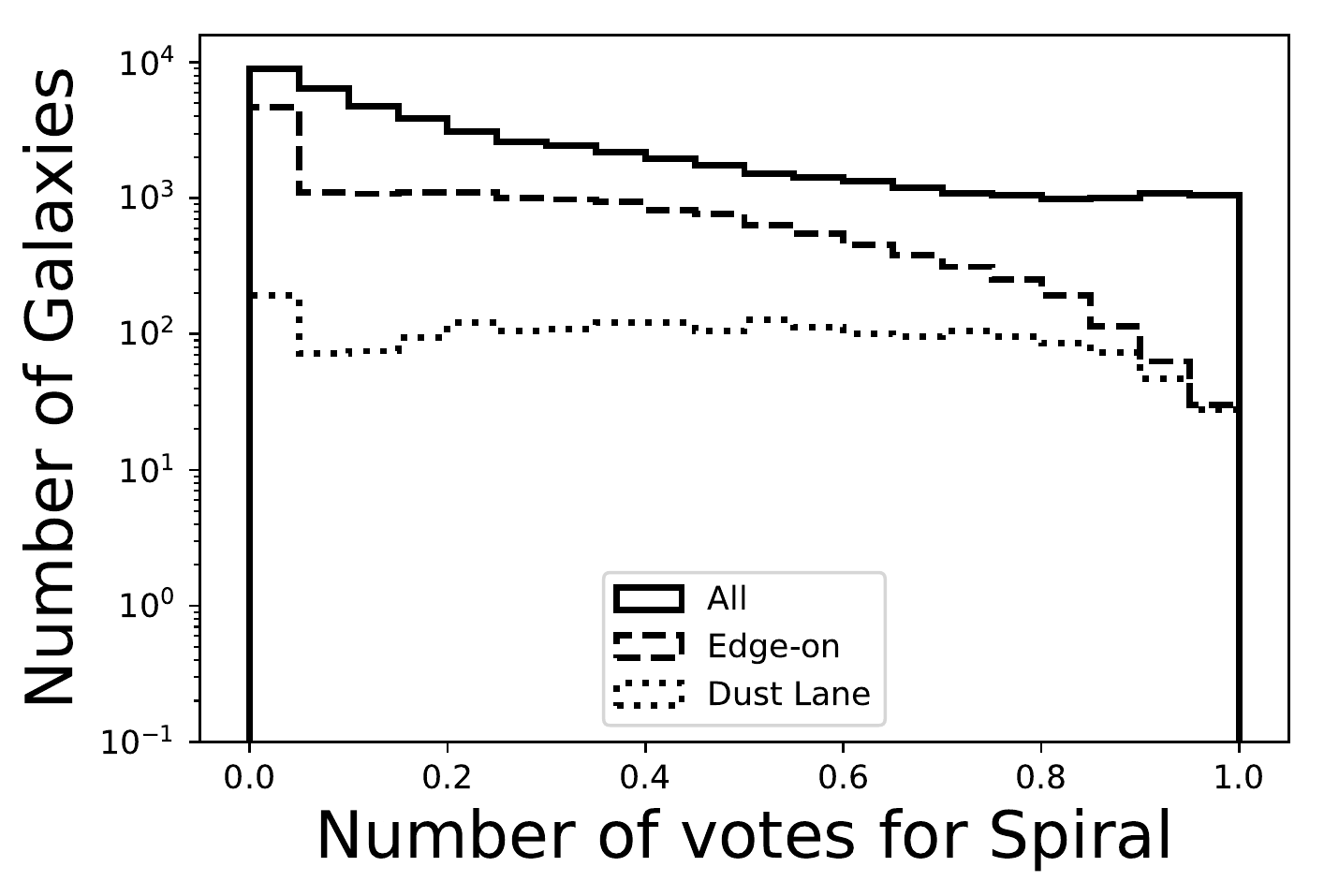}
\caption{\label{f:votes}The number GAMA/KiDS galaxies as a function of the fraction of votes in favor of galaxy with features (A01 in Figure \ref{f:tree}), edge-on (T01 in Figure \ref{f:tree}), and a dustlane (T06 in Figure \ref{f:tree}). 
The solid line are all the objects and the fraction of votes in favor of a galaxy with features, the dashed line is a subset of these ($\rm f_{features} > 0.5$) and the fraction of votes in favor of the galaxy being edge-on. The dotted line is a subset of these ($\rm f_{features} > 0.5$ and $\rm f_{edge-on} > 0.5$) with the fraction of votes in favor of a dust lane.}
\end{figure}

\begin{figure*}
\includegraphics[width=\textwidth]{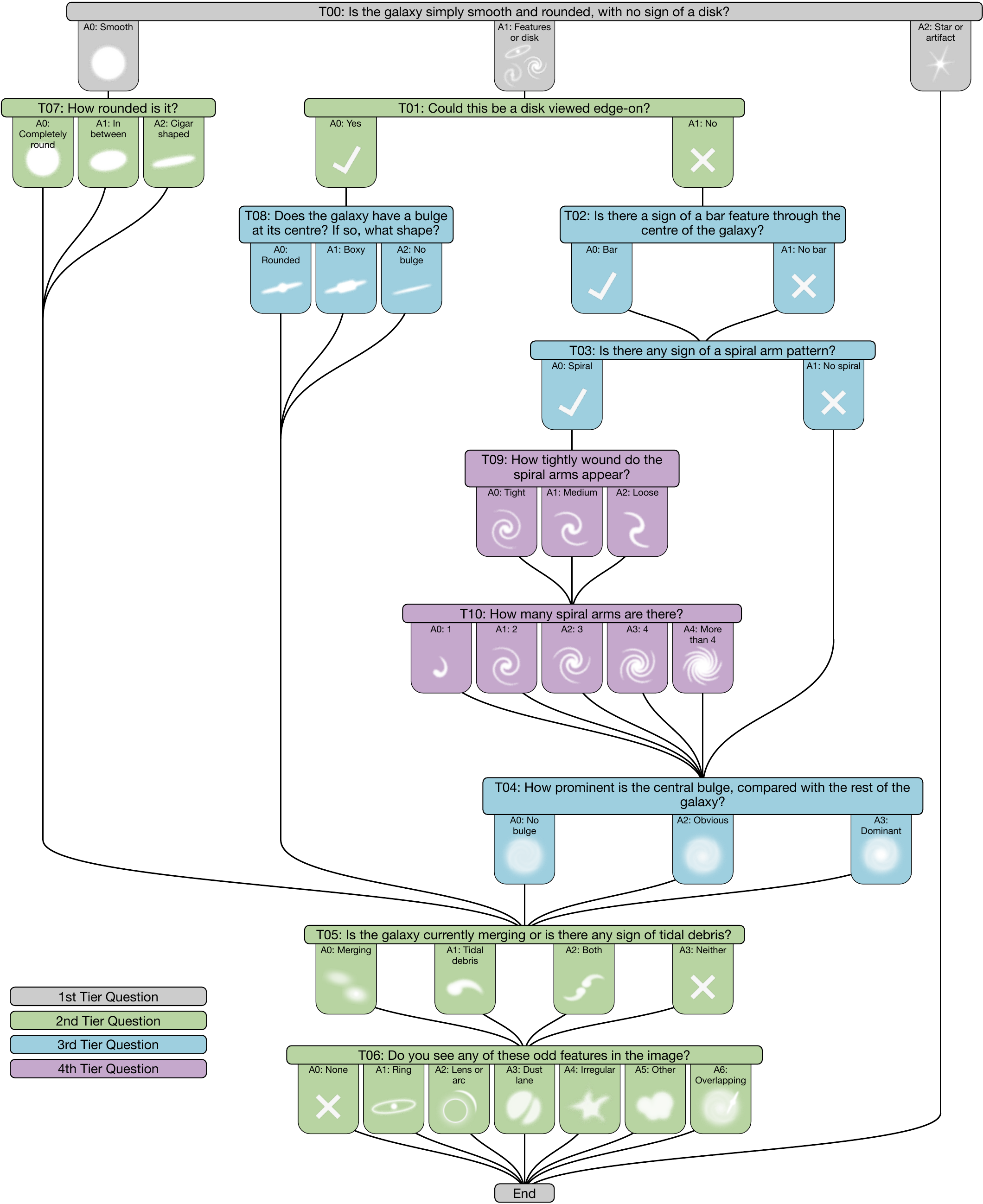}
\caption{\label{f:tree}The decision tree of the Galaxy Zoo iteration 4, which was followed by the GAMA/KiDS GZ iteration.}
\end{figure*}

GAMA is a combined spectroscopic and multi-wavelength imaging survey designed to study spatial structure in the nearby ($z < 0.25$) Universe on kpc to Mpc scales \citep[see][for an overview]{Driver09, Driver11}. The survey, after completion of phase 2 \citep{Liske15}, consists of three equatorial regions each spanning 5 deg in Dec and 12 deg in RA, centered in RA at approximately 9h (G09), 12h (G12) and 14.5h (G15) and two Southern fields, at 05h (G05) and 23h (G23). The three equatorial regions, amounting to a total sky area of 180 deg$^2$, were selected for this study. For the purpose of visual classification, 49851 galaxies were selected from the equatorial fields with redshifts $z<0.15$. Figure \ref{f:votes} shows the distribution of votes for galaxies in our subsample of disk galaxies (T00 in Figure \ref{f:tree} question has been answered by more than 50\% of the volunteers as a disk galaxy).  The GAMA survey is $>$98\% redshift complete to r $<$ 19.8 mag in all three equatorial regions. We use two data-products described in the third GAMA data-release \citep[DR3,][]{Baldry18}: the {\sc magphys} SED fits \citep{Driver18} and the S\`ersic fit catalogs \citep{Kelvin14}. 


The GAMA-KiDS Galaxy Zoo project uses the decision tree in use for the latest (4th) iterations of the Zoo. KiDS cutouts were introduced to the classification pool and mixed in with the ongoing classification efforts. Scientific aims include correlating general morphology to the GAMA results using the full suite of multi-wavelength and spectral information and the identification of rare features (e.g. strong lensing arcs of galaxy occultation). 
A full description of the GAMA-KiDS Galaxy Zoo effort can be found in Kelvin et al. {\em in preparation}.

In addition to the GAMA-KiDS Galaxy Zoo classifications, we use the {\sc magphys} \citep{magphys}, spectral energy distribution fits to the GAMA multiwavelength photometry \citep{Wright17}, presented in \cite{Driver18}. {\sc magphys} computes stellar mass, specific star-formation rate, dust mass, cold dust fraction, cold dust temperature, average face-on optical depth of each galaxy, which we can compare against the dust lane identifications in the GAMA-KiDS Galaxy Zoo data.
%
In addition to the {\sc magphys} data, use the  S\`ersic fits to the UKIDSS \citep{Kelvin14}.

Figure \ref{f:sm_ssfr:z} shows the {\sc magphys} stellar mass and specific star-formation rate plot with the redshift indicated as well. One can discern selection effects e.g. how lower-mass objects are only found at lower redshifts and more massive objects can be found out to the highest redshifts.


\begin{figure}
\includegraphics[width=0.5\textwidth]{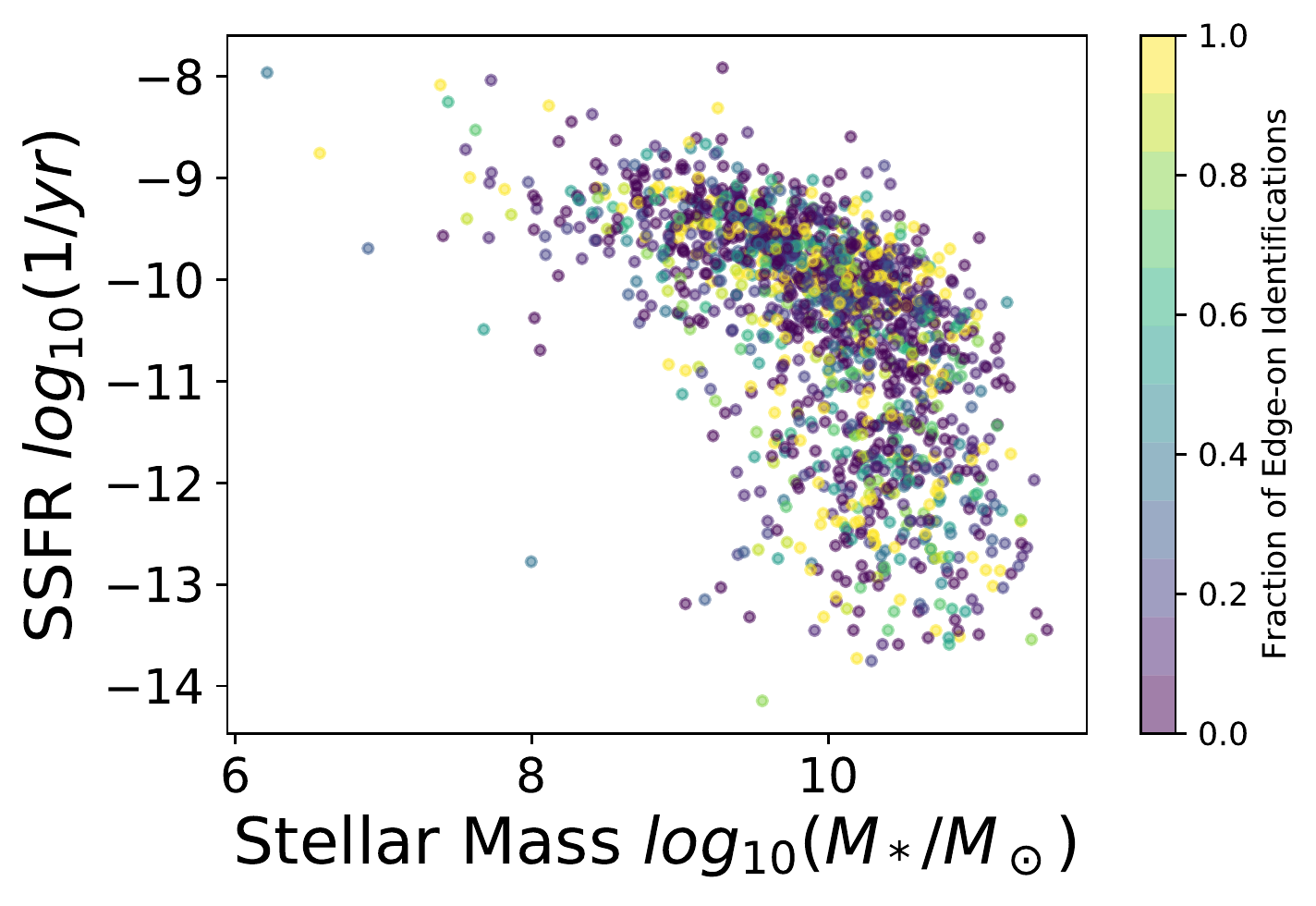}
\caption{\label{f:sm_ssfr:edgeon} The relation between stellar mass from the {\sc magphys} fit,  and star-formation with the fraction of votes in favor of an edge-on disk. Galaxies voted in favor of edge-on disk are spread throughout star-formation rate and stellar mass. }
\end{figure}

\section{Galaxy Zoo Decision Tree} \label{s:gz}

Figure \ref{f:tree} shows the decision tree for the Galaxy Zoo fourth iteration, of which the GAMA/KiDS classifications are part. After the initial decision if a galaxy is smooth, disk, or star/artifact (T00), second tier questions are asked of the volunteers (T07 or T01). In the case of disk galaxies, the follow up question determines if the disk is viewed edge-on or not (T01). This is a critical question for this project as we are interested in the prevalence of dust lanes in edge-on disks. The next follow up is to determine what kind of bulge is visible: none, a box-shaped or round one (T08). The next tier question is whether any signs of interaction with a nearby galaxy is evident (T05). Finally, the last question is a series of morphological features (ring, lens arc, dust lane, irregular, other or overlap). More than one choice can be marked. 
Figure \ref{f:votes} shows the distribution of votes in favor of disk galaxies (T00), edge-on (T01), and displaying a dust lanes (T06). 
To select a galaxy as having a feature, we require 50\% of the votes in favor of a disk or edge-on ($f_{\rm disk}>$50\% or $f_{\rm edgeon}>$50\%) and 10\% for the identification of a dust lane ($f_{\rm dustlane}>$10\%) because votes for them are rare (Figure \ref{f:votes}). The selection threshold for dustlanes is set to be inclusive because other morphological features are rarely mistaken for dust lanes and dust lanes are so commonly thought of as an edge-on ``normal'' feature that they are not remarked upon. Voting fractions have been de-biased using the now standard Galaxy Zoo calibrations of votes \citep[see Kelvin et al. {\em in preparation}][]{Hart16}. 

The improvement over the original Galaxy Zoo is that these questions are not behind a gate question of ``is there anything odd?". Many users considered dust lanes not odd and would therefore not choose the ``odd" button. 
The remaining issue is that votes for one morphological features may draw away votes from another. Nevertheless, we assume all these features are relatively rare enough for this to be not too great an issue. 

\subsection{Edge-on Disk Identification} \label{s:edgeid}

We select edge-ons by requiring half of the votes by the volunteers in favor of the edge-on question. Figure \ref{f:sm_ssfr:edgeon} shows the number of edge-on votes in the stellar mass and specific star-formation plot. Dust lanes in thicker edge-on disks (more massive galaxies) can be identified out to greater distances (Figure \ref{f:sm_ssfr:z}). There is therefore an unavoidable bias in our sample against more distant, low-mass galaxies, both in the GAMA/KiDS Galaxy Zoo survey and the edge-on identification in the KiDS images.

We opted for Galaxy Zoo identification of edge-on disks even though it allows for many more possible disk inclinations than other selection methods (e.g. near-infrared ellipticity) because we wanted to compare the results to these galaxy properties. For example, we want to know the effects of a substantial bulge and an ellipticity selection, as has been typically done before, would bias against early (Sa etc.) types.

The term ``edge-on" is somewhat subjective. In Galaxy Zoo 2, using SDSS images, about 20\% of \cite{Willett13} of disk galaxies are considered ``edge-on", using a 70\% votes in favor. Here we use a looser fraction (50\%) but this may well affect the final fraction of galaxies identified with a dust lane as well.

\begin{figure}[h]
\includegraphics[width=0.5\textwidth]{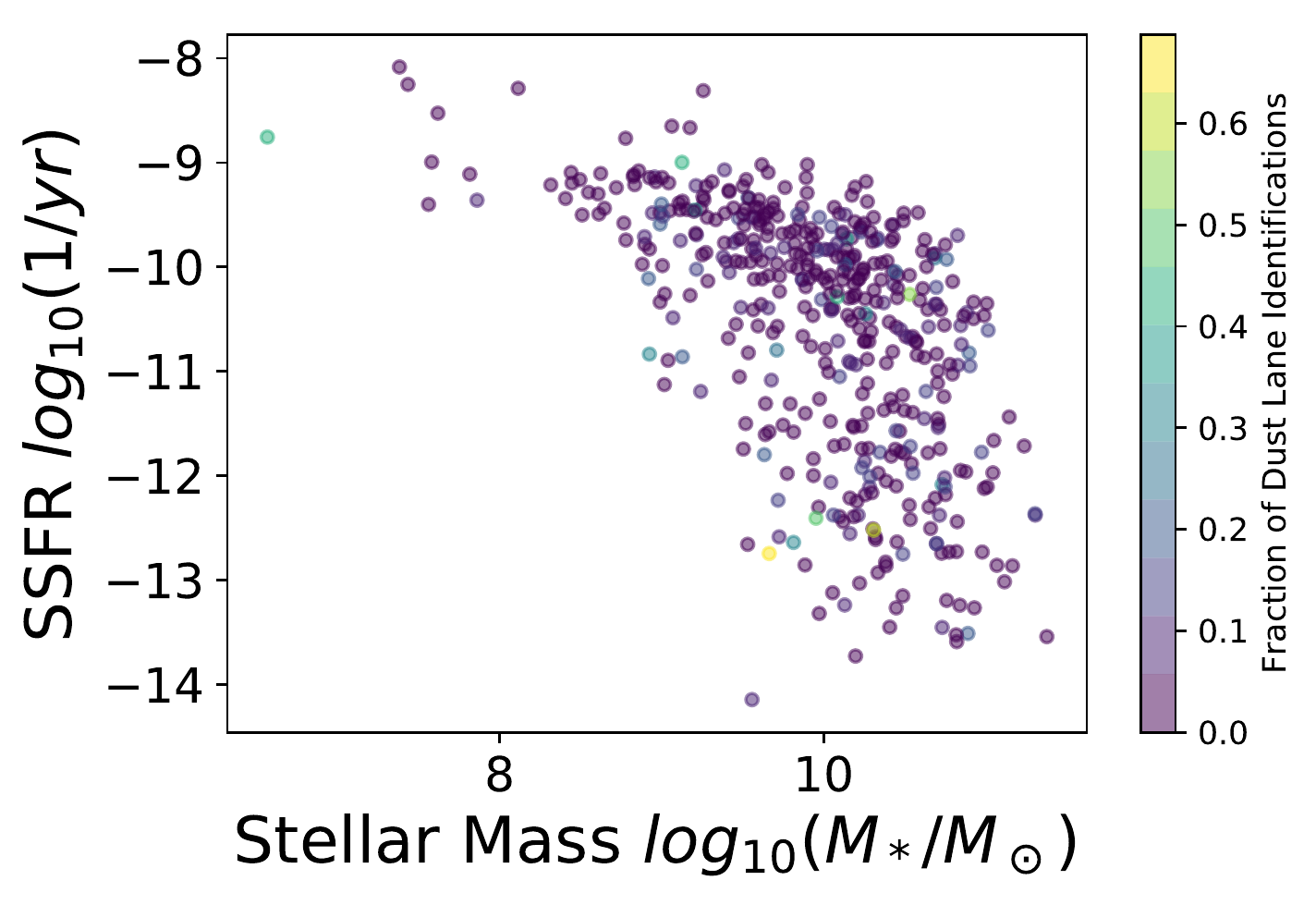}
\caption{\label{f:sm_ssfr:dl} The relation between stellar mass and star-formation from the {\sc magphys} fit, with the fraction of votes identifying a dust lane in galaxies with $>50$\% of the vote in favor of edge-on disk. }
\end{figure}
\subsection{Dust Lane Identification} \label{s:dustlaneid}

\begin{figure*}
\includegraphics[width=0.195\textwidth]{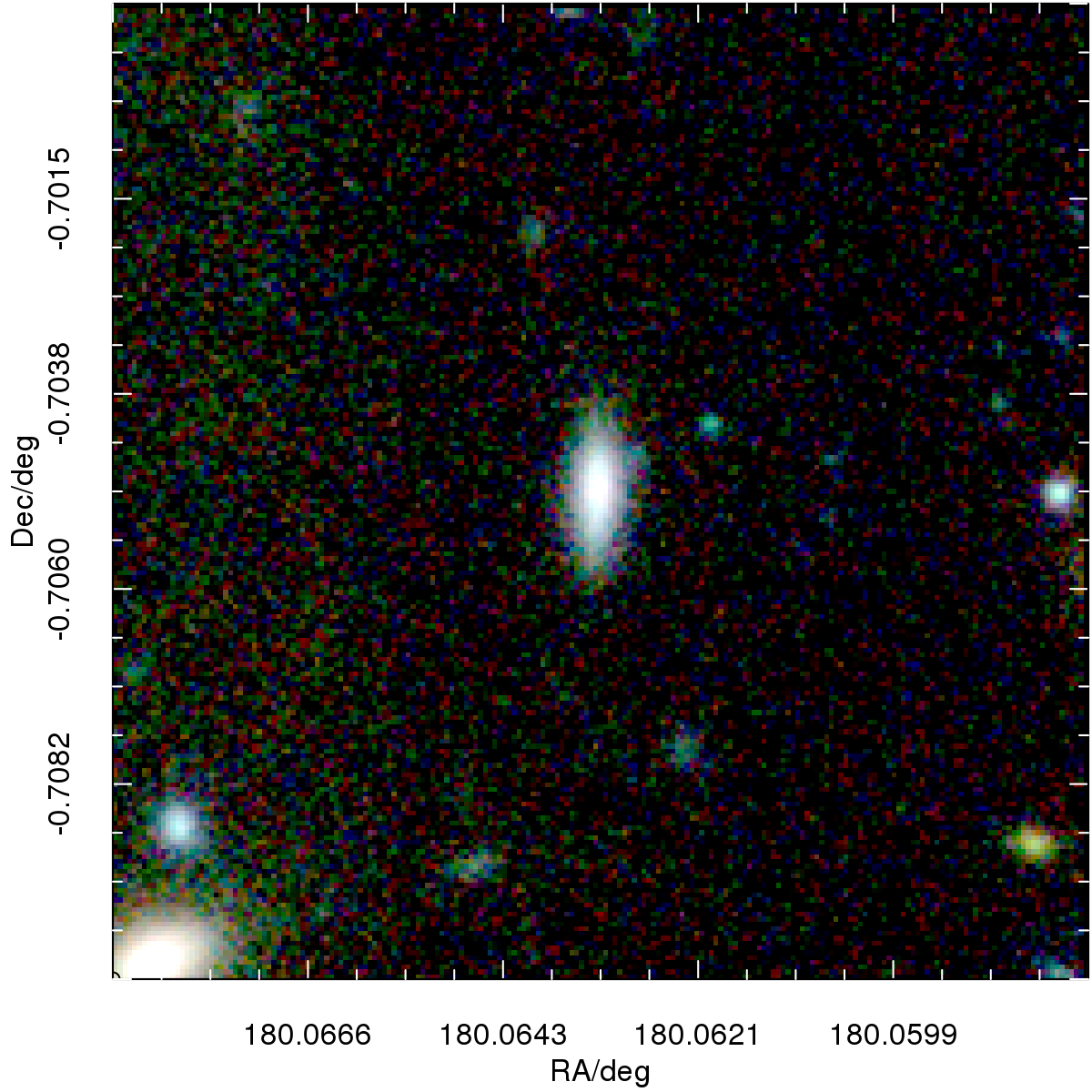}
\includegraphics[width=0.195\textwidth]{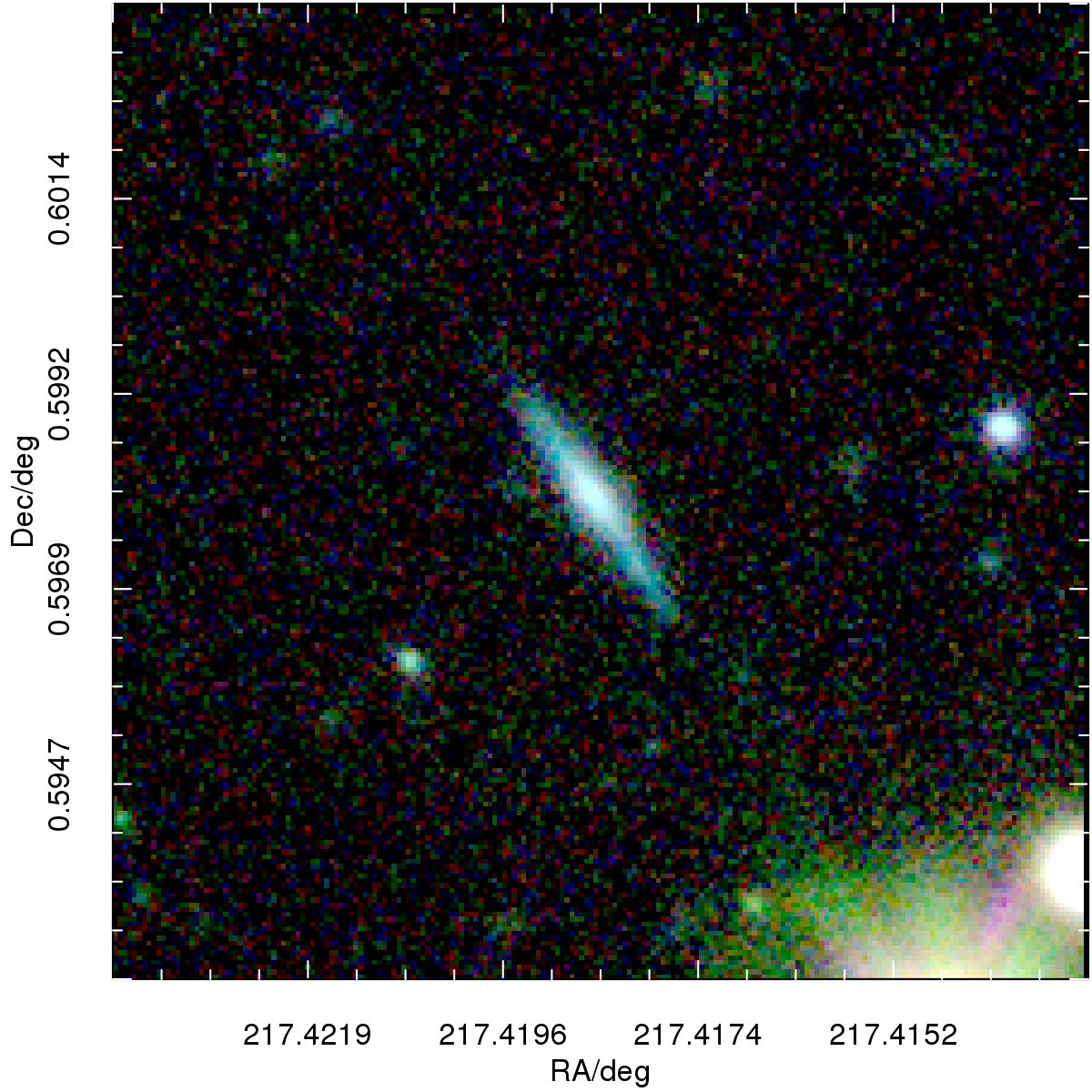}
\includegraphics[width=0.195\textwidth]{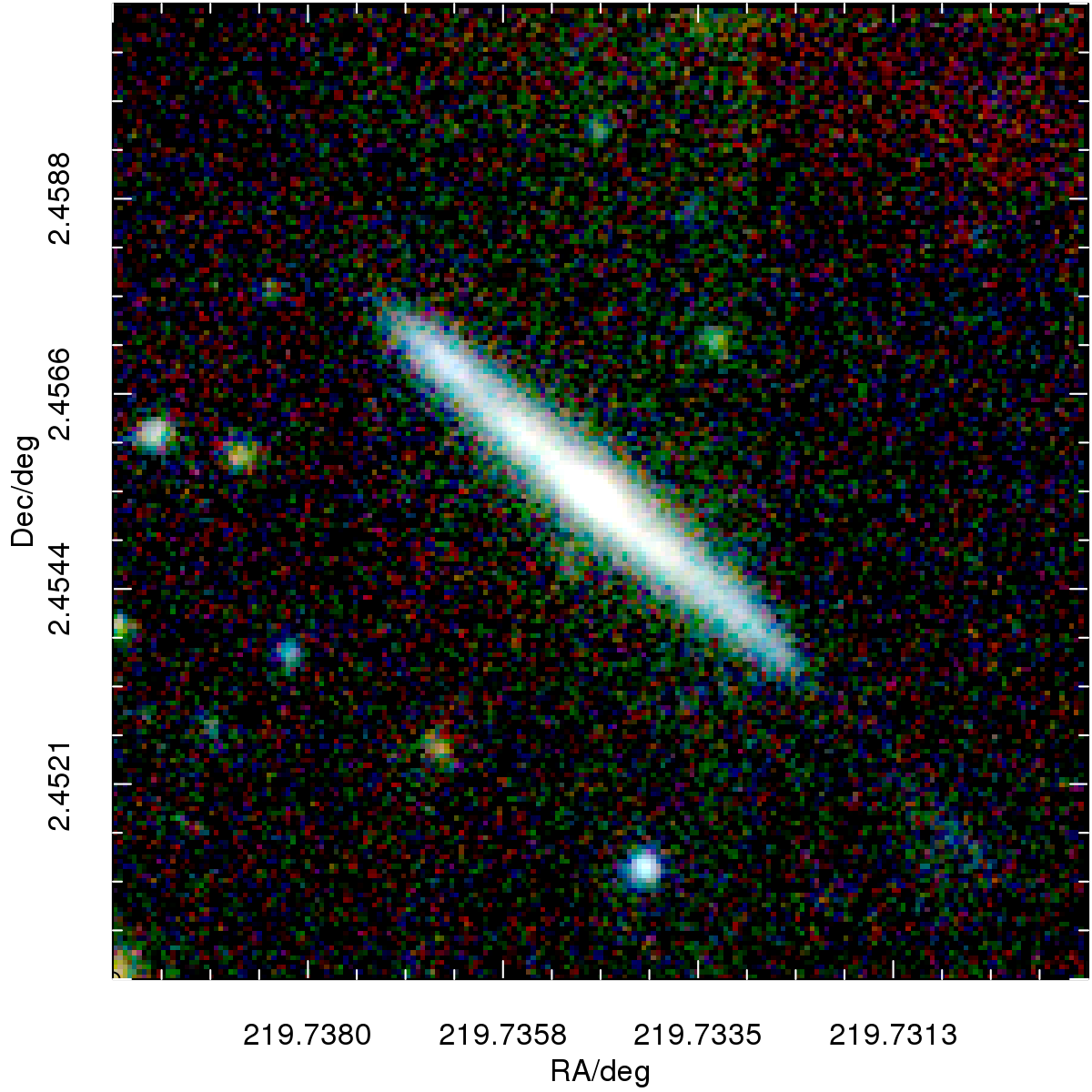}
\includegraphics[width=0.195\textwidth]{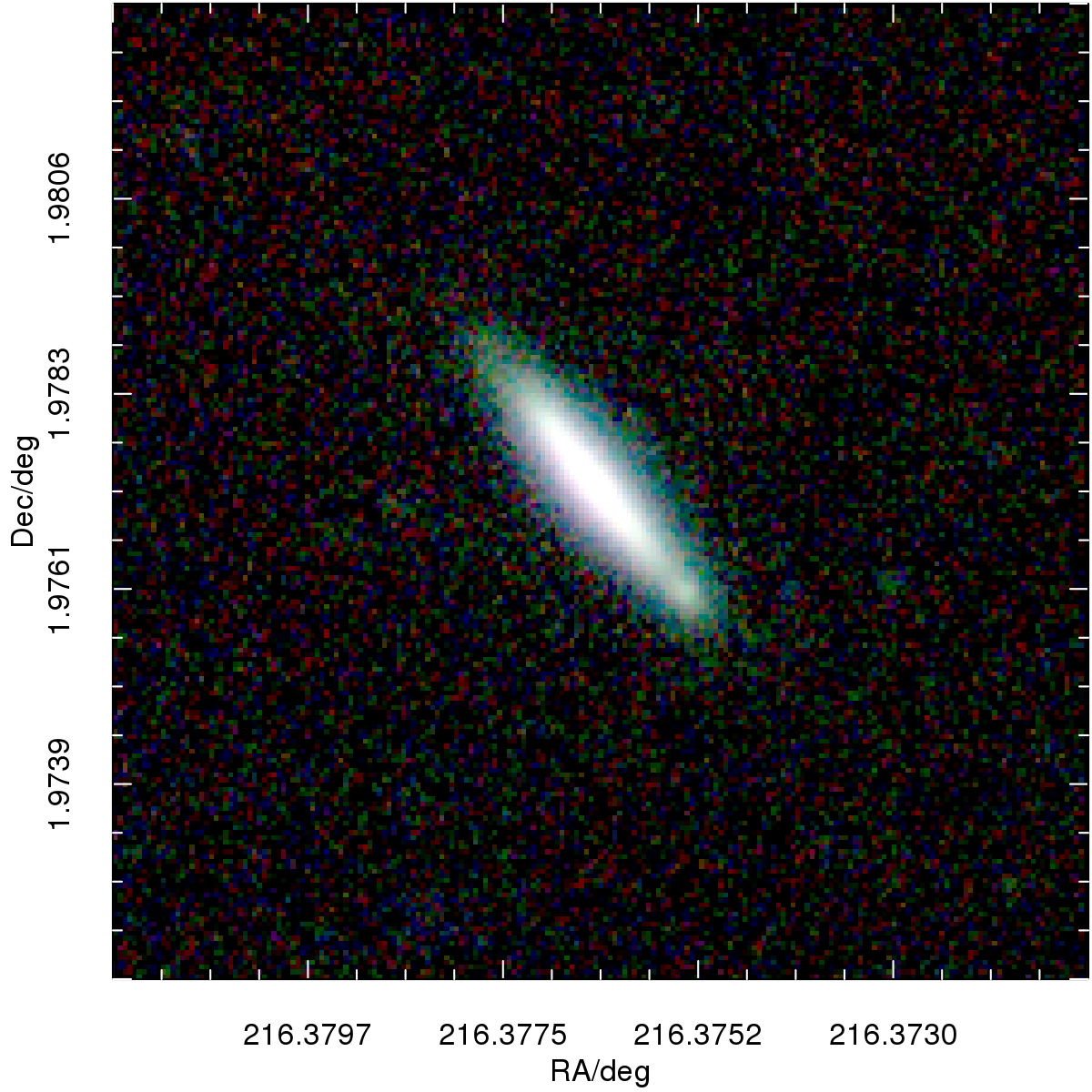}
\includegraphics[width=0.195\textwidth]{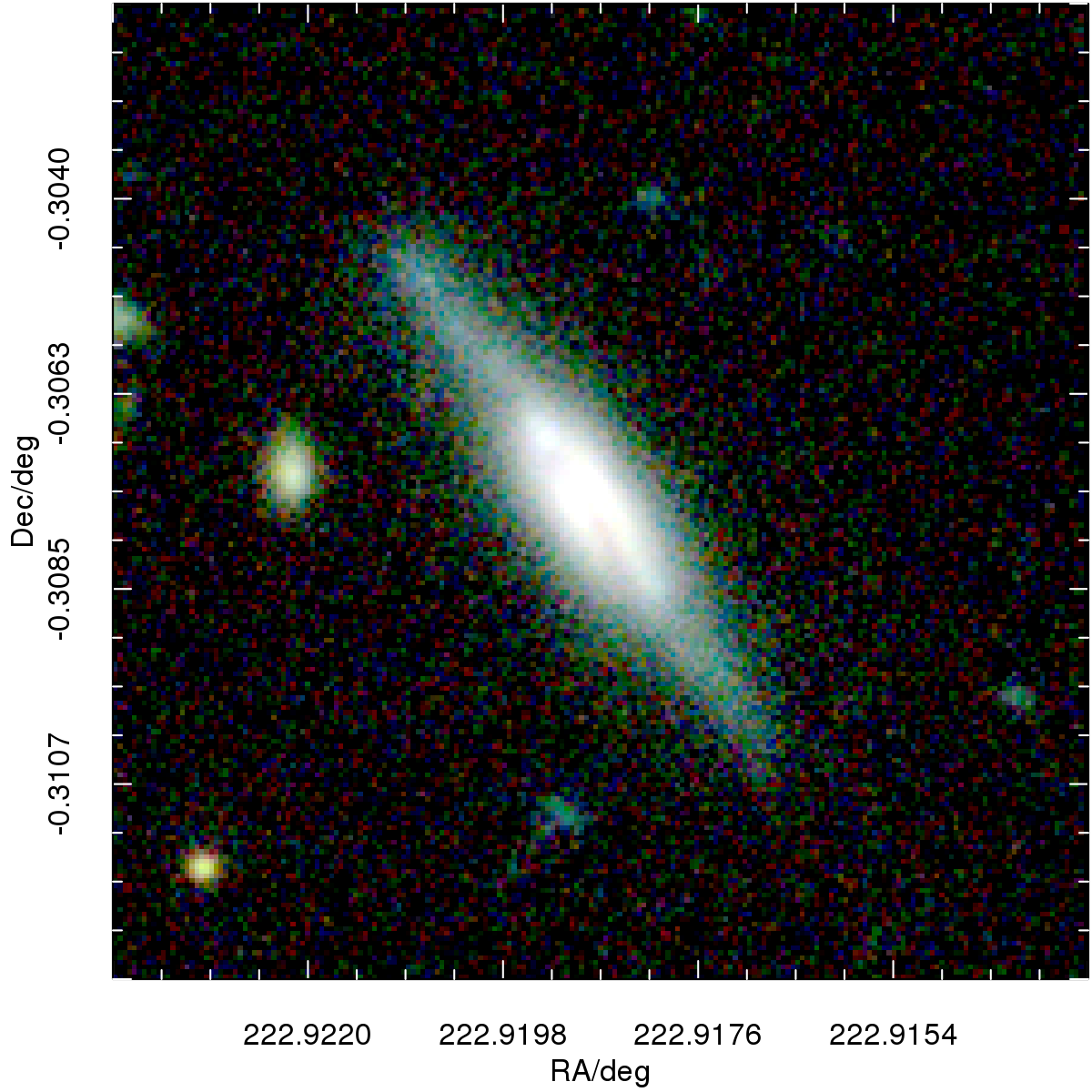}
\includegraphics[width=0.195\textwidth]{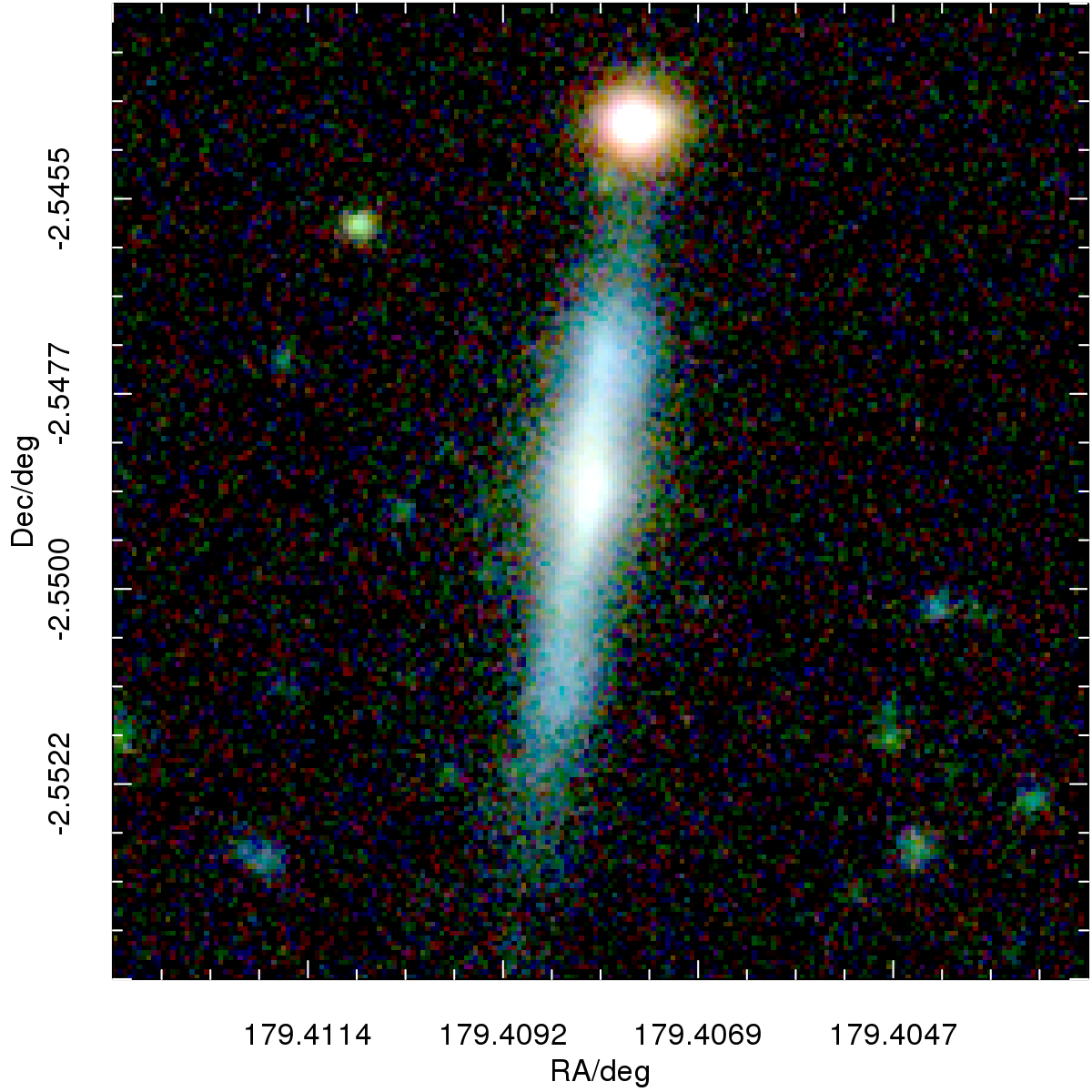}
\includegraphics[width=0.195\textwidth]{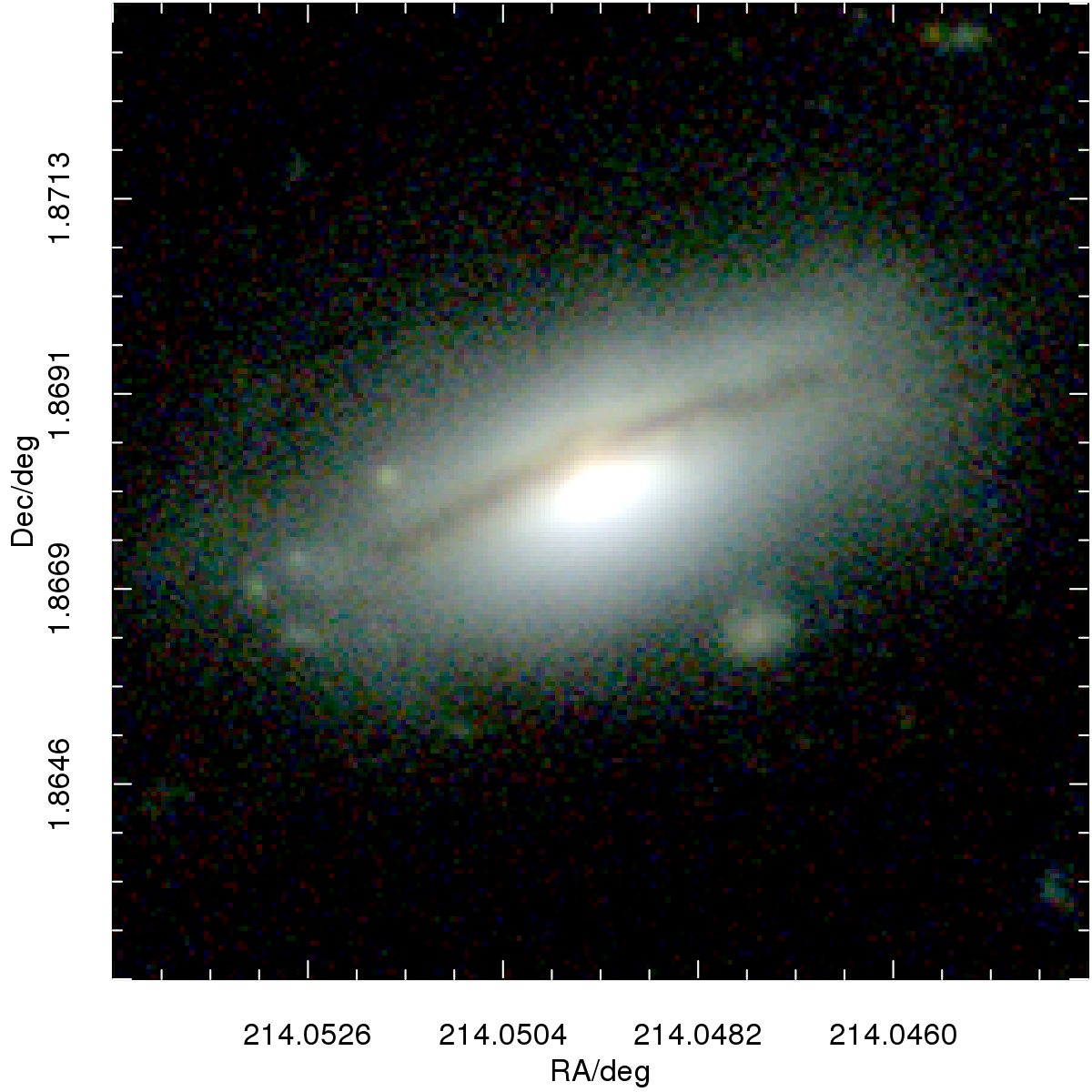}
\includegraphics[width=0.195\textwidth]{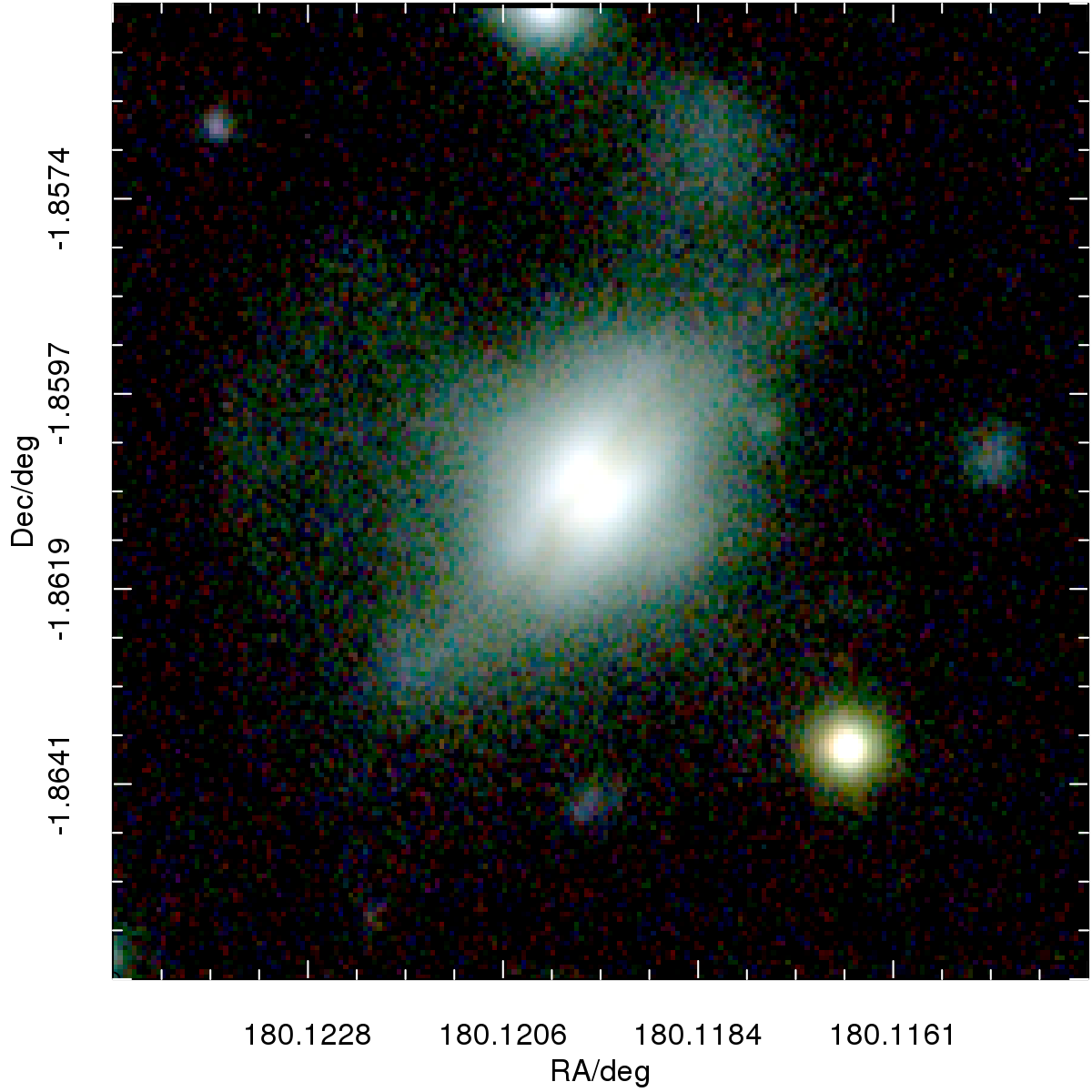}
\includegraphics[width=0.195\textwidth]{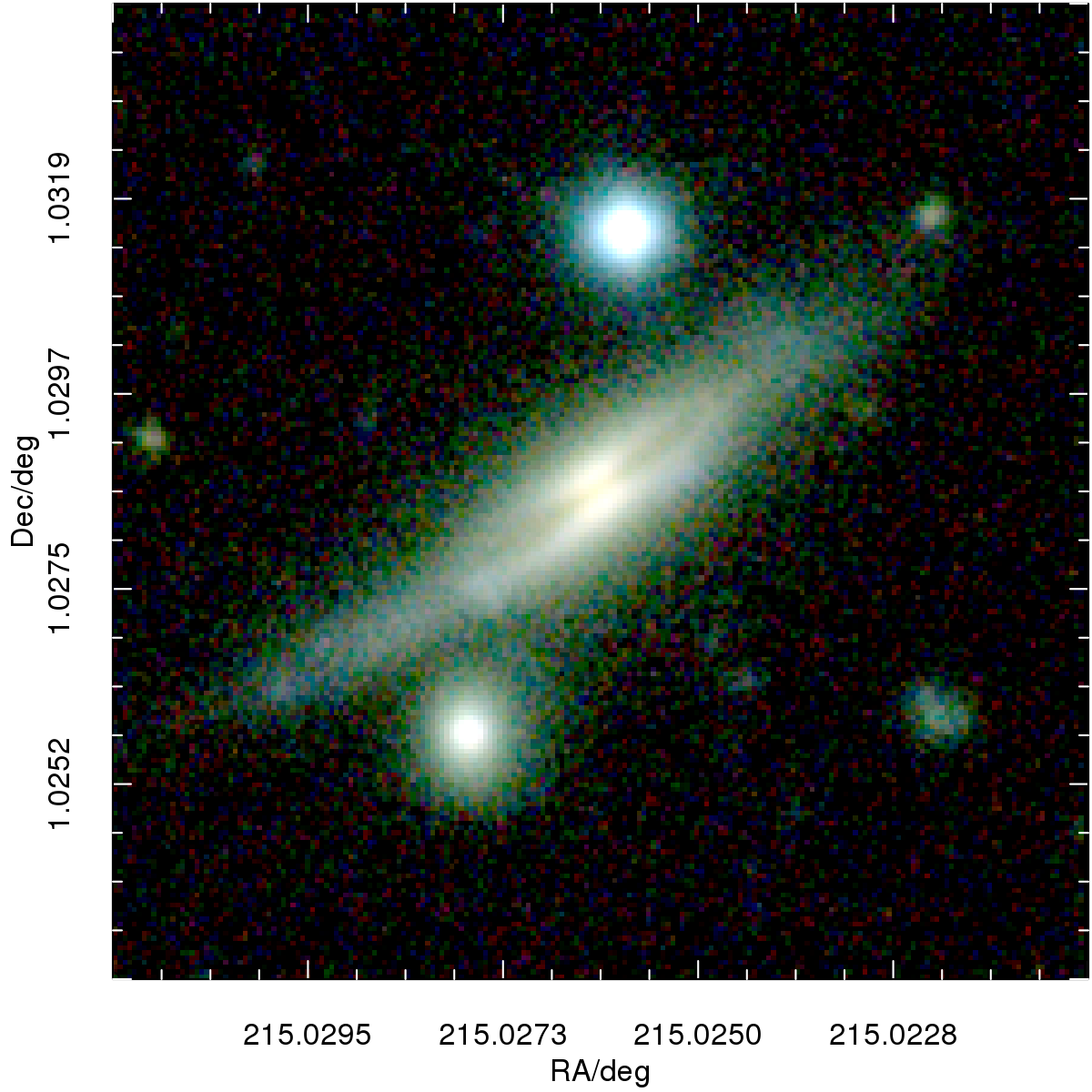}
\includegraphics[width=0.195\textwidth]{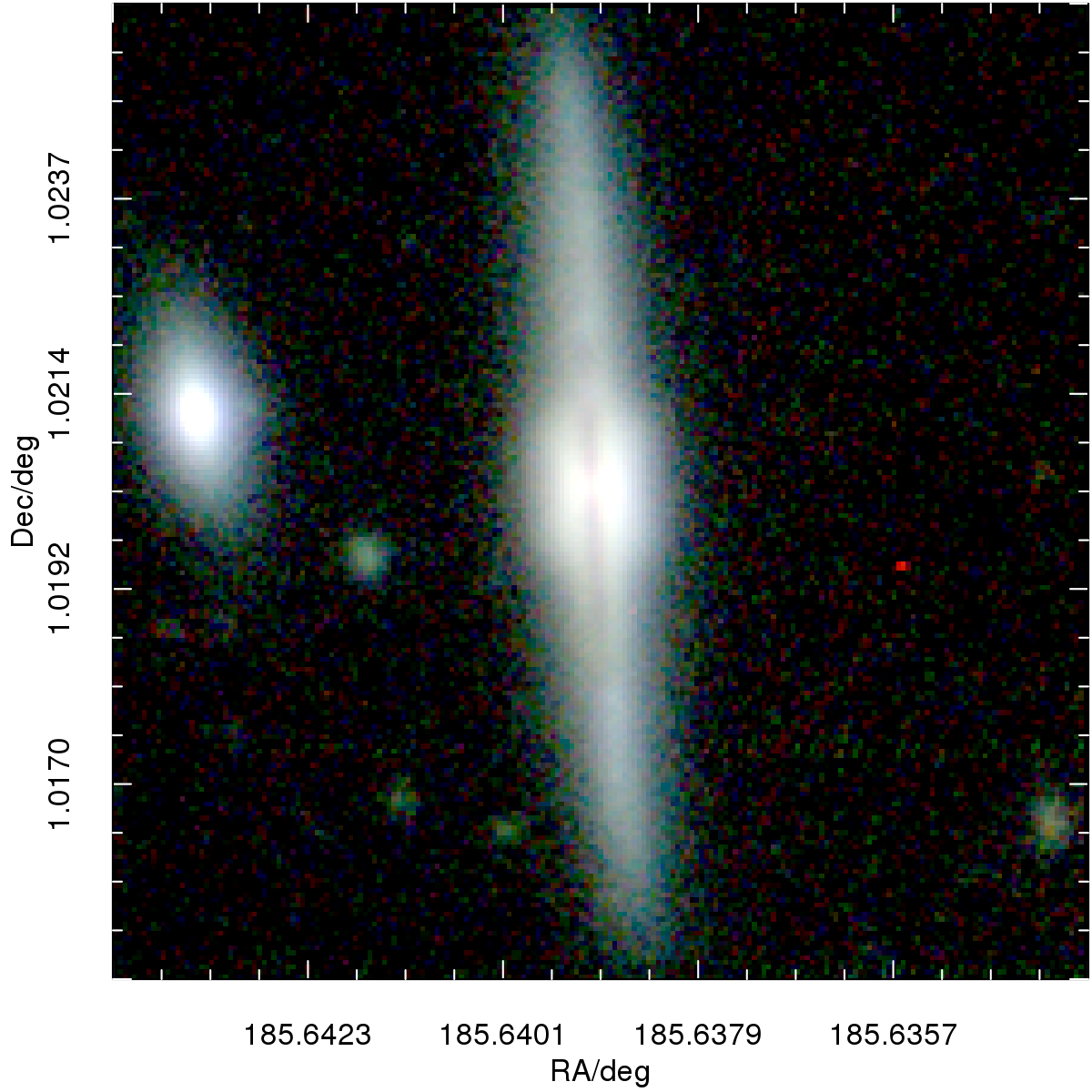}
\caption{\label{f:kids:examples} Randomly selected examples with $f_{edgeon} > 0.5$ from the GalaxyZoo GAMA sample in order of dust lane vote fraction. Starting at the top left with $f_{dustlane}=0$ and ending with $f_{dustlane}=0.9$bottom right. }
\end{figure*}

Figure \ref{f:sm_ssfr:dl} shows the fraction of dust-lane votes in the stellar mass and specific star-formation plot. 
Dust lanes votes occur throughout the stellar mass and specific star-formation. Because there are many options to choose from in the final question, we only require 10\% of the votes for us to consider the galaxy to have a dust lane. Requiring a higher fraction leads to similar results but with lower statistical confidence due to a smaller sample. Our reasoning is that dust lanes are not very remarkable so a few votes in favor means it is clearly present. Figure \ref{f:kids:examples} shows a few randomly drawn examples of the edge-on sample with different fractions of votes in favor of a dust lane. We caution against using examples such as these to draw a criterion; Figure \ref{f:kids:examples} is purely for illustrative purposes. 

At the maximum distance of z=0.14, the KiDS nominal resolution (0\farcs6) corresponds to $\sim1.6$ kpc. Only in the most massive disk systems would a dust lane stand out at this resolution (e.g. NGC 891). The GAMA/KiDS Galaxy Zoo sample is less complete at the low-mass end with a maximum redshift for the low-mass ($M^*< 10^9 M_\odot$) galaxies at $z\sim0.04$, corresponding to a linear resolution of half a kiloparsec, enough to identify a dust lane.

Dust lane identification is therefore incomplete at the high-mass end (part of the sample is too far away for positive identification in all cases) and incomplete at the low-mass end due to survey volume. Over the entire sample of edge-on galaxies (identified as such by the Galaxy Zoo) about 50 per cent of the edge-on galaxies display a dust lane. \cite{Dalcanton04,Obric06} and \cite{Holwerda12a} find that the dust lane fraction lies around 80\% for massive disks. 

We note that there are inevitable biases introduced by the Galaxy Zoo voted selection for edge-ons: inclusion of much more earlier type galaxies, not necessarily disk dominated ones and galaxies not perfectly edge-on. These will inevitably lower the overall number of galaxies with a dust lane overall \citep[see the discussion in ][]{Kaviraj12a}.

\section{Results} \label{s:results}

We plot the fraction of the galaxies along a property (e.g. stellar mass or dust temperature) with the full disk galaxy sample, those considered edge-on, and finally those considered edge-on with a dust lane identified.
Uncertainties are calculated from the parent sample and the fraction identified using the prescription in \cite{Cameron11}. 
%
%

\subsection{{\sc magphys} Stellar Mass and (Specific) Star-formation Rate} \label{s:magphys:star}

\begin{figure}
\includegraphics[width=0.5\textwidth]{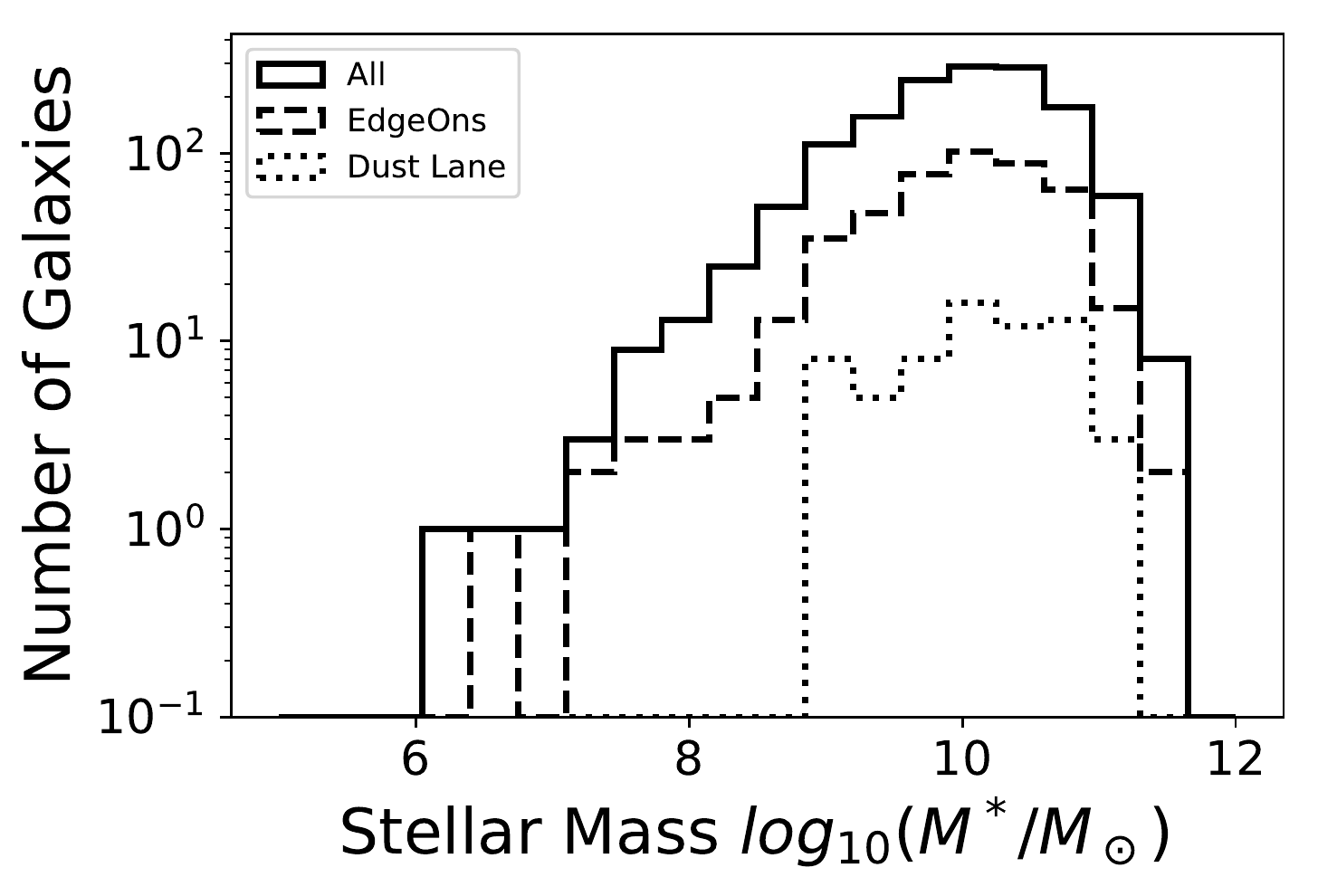}
\includegraphics[width=0.5\textwidth]{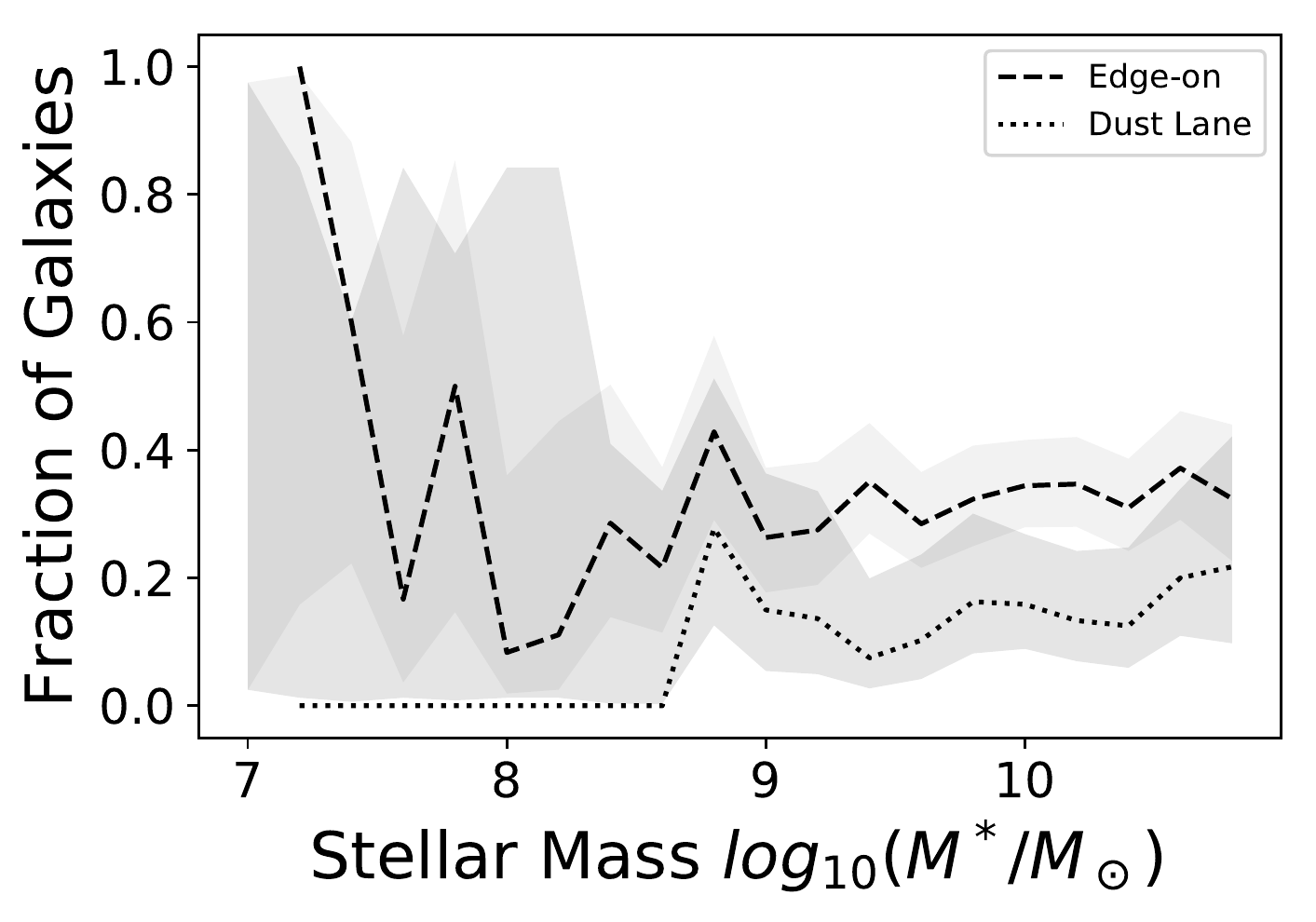}
\caption{\label{f:hist:mstar} The histogram (top) of the stellar mass of all the galaxies voted as disks (solid line), edge-on (dashed line) and with a dust lane (dotted line). The fraction of galaxies classified as edge-on (dashed line, $f_{\rm edgeon} > $50\% of the votes for selection) and the fraction that is voted both edge-on and displaying a dustlane (dotted line, $f_{\rm edgeon} > 50$\%, $f_{\rm dustlane} > 10$\%). Gray shaded areas are the standard deviation in the numbers in each bin, both edge-on votes and edge-on and with a dust lane. Dust lanes are identified in galaxies more massive than $10^{8.5} M_\odot$, consistent with previous dust lane searches in the local and distant Universe \protect\citep{Dalcanton04,Holwerda12a}. 
We will use fractions of the galaxy populations for further comparisons. }
\end{figure}

\subsubsection{Stellar Mass} \label{s:mstar}

Figure \ref{f:hist:mstar} shows the distribution of stellar mass, as determined by {\sc magphys} in our sample, both as a histogram and a fraction of all the disk-identified galaxies with errors calculated using the prescription from \cite{Cameron11}.
Edge-on disk galaxies follow the full sample of disk galaxies very well. We note a complete absence of edge-on galaxies with a dust lane below $10^9 M_\odot$. Galaxies with masses below this limit are only included in the GAMA/Galaxy Zoo with redshifts below $z\sim0.04$. On a linear scale, the KiDS resolution (0\farcs6) translates approximately 0.5 kpc at this distance. Thus any clear dustlanes in the low-mass galaxies should be identifiable. 

Several selection effects in the identification of both edge-ons and dust lanes may well play a role here. 
The GAMA survey is the complete for the lower-mass edge-ons in the smallest volume. The small number statistics at the lower end means that if a few low-mass edge-on galaxies with a dust lane have been misidentified, a similar fraction as the higher mass bins could still be true.
The Galaxy Zoo identification scheme may well classify lower-mass edge-on spirals as "smooth" as several examples of this class show few distinguishing features, which includes features such as dust lanes. In this latter scenario, the number of low-mass edge-on disks would go up but  unlikely that the fraction of low-mass edge-on galaxies with a dust lane would increase.

The trend in Figure \ref{f:hist:mstar} with mass is consistent with the result from \cite{Dalcanton04}, who noted a distinct changeover below and above 120 km/s rotation speed ($M^* \sim 10^{9.8} M_\odot$). \cite{Dalcanton04} and \cite{Holwerda12a} find that 80\% of the massive disks have a dust lane. In this Galaxy Zoo sample, the fraction lies lower however.

The fraction of galaxies identified with a dust lane is approximately half of those identified as edge-on but consistent with the fraction of 80\% found by previous authors. 




\begin{figure}
\includegraphics[width=0.5\textwidth]{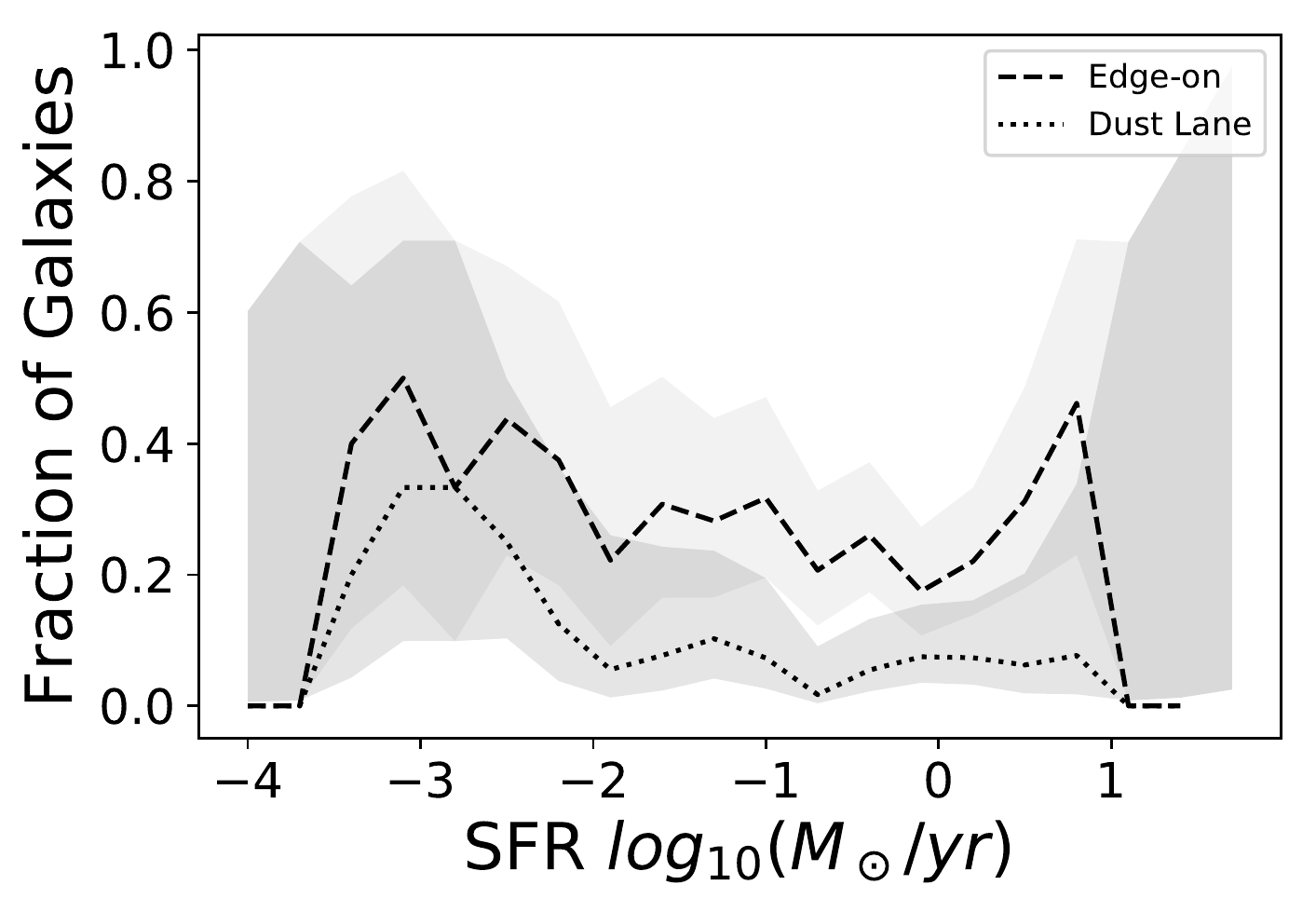}
\caption{\label{f:hist:sfr} The fraction of galaxies marked as edge-on (dashed lines) or edge-on with a dust lane (dotted line) as a function of the star-formation rate of all the galaxies. Gray shaded areas are the uncertainties in both fractions.}
\end{figure}
\subsubsection{Total Star-Formation Rate} \label{s:sfr}

Figure \ref{f:hist:sfr} shows the fractions of galaxies as a function of star-formation for edge-on and dust lane identification. At any given level of star-formation, dust lanes occur at the same fraction in edge-on galaxies. Naively, one could expect the star-formation rate influencing or being influenced by the presence of a dust lane. For example, a dark dust lane, associated with a compact molecular component of the ISM that fuels star-formation or the additional turbulence thanks to newly formed stars could dissipate the compact dust lane. But no dependency on the total star-formation rate for the prevalence of dust lanes is evident. Models for massive edge-on disks \citep{Popescu00,Popescu11,Popesso12} find that the edge-on dust lane is more the line-of-sight effect of the diffuse component of the dusty ISM rather than associated with the star-forming dust clouds.

\begin{figure}
\includegraphics[width=0.5\textwidth]{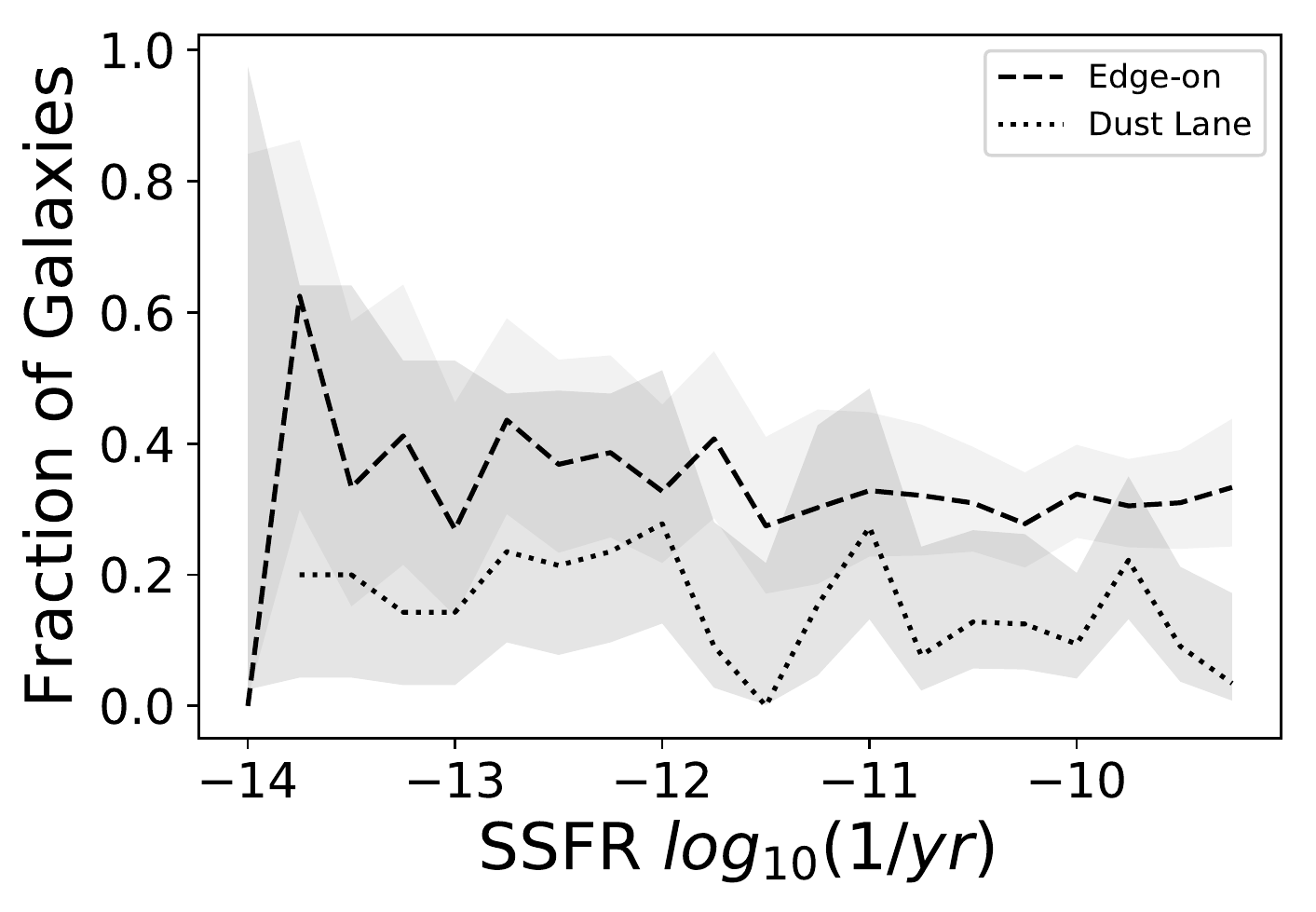}
\caption{\label{f:hist:ssfr} The fraction of galaxies as a function of the specific star-formation rate voted as edge-on (dashed line) and with a dust lane (dotted line). Gray shaded areas are the uncertainties in both fractions.}
\end{figure}

\subsubsection{Specific Star-Formation Rate} \label{s:sfr}

Figure \ref{f:hist:ssfr} shows the histogram for specific star-formation for edge-ons and those with dust lanes. At any given level of specific star-formation, dust lanes occur at the same fraction in edge-on galaxies. Specific star-formation is the relative growth of the stellar population and we hoped for a better indicator of what is the dominant mechanism rearranging the dusty ISM: gravitational contraction balanced by turbulence dispersing the molecular clouds throughout the height of the disk. However, like star-formation, there is no clear specific star-formation rate where dust lanes become more prevalent in edge-on galaxies.

\subsection{{\sc magphys} Dust Output Parameters} \label{s:magphys:dust}

{\sc Magphys} outputs several parameters directly related to the dusty ISM of a galaxy as it reprocesses star-light into far-infrared and sub-mm emission. {\sc magphys} treats the dusty ISM as a diffuse disk of colder ISM with clumps of heated dust close to the ongoing star-formation. 
{\sc Magphys} output includes the dust mass, the fraction of dust mass in the cold component, the temperature of the cold component, and the average face-on optical depth in V-band ($\tau_V$). {\sc Magphys} was calibrated on local galaxies and the edge-on perspective on disk galaxies is an edge-case to test it on, with much greater fraction of the ISM along the line-of-sight effectively opaque. Therefore, the infrared emission from the dust in the dense dust lane would be wrongly associated to the optically thin dust at larger vertical scales. Thus the {\sc magphys} output may be biased due to this mismatch in emission and attenuation effects.

The total dust mass or the ratio of stellar to dust mass are prime candidates for {\sc magphys} output to correlate with the presence of a dust lane in edge-on galaxies. One would naively expect for example that more dust mass or relatively more dust mass would increase the likelihood of a dust lane if dust is distributed relatively evenly (diffuse component) throughout a disk galaxy.

\begin{figure}
\includegraphics[width=0.5\textwidth]{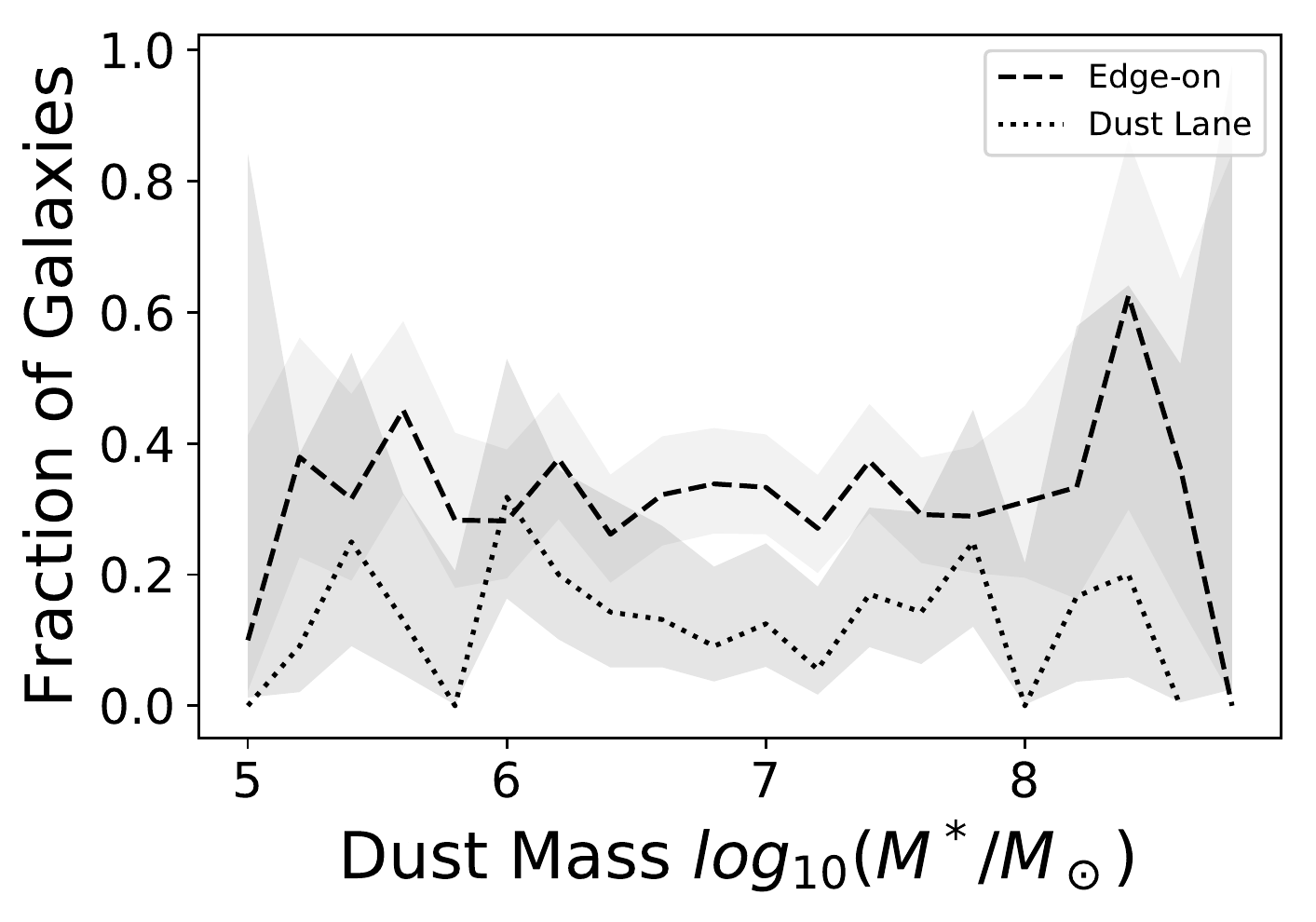}
\caption{\label{f:Md} The fraction of disk galaxies voted edge-on (dashed line) and those voted disk, edge-on and showing a dust lane (dotted line) as a function of {\sc magphys} dust mass. Dust lanes are identified throughout the range of {\sc magphys} dust masses $M_D \sim 10^{5-6} M_\odot$. }
\end{figure}

\subsubsection{Dust Mass} \label{s:Md}

{\sc magphys} reports a total dust mass for each galaxy. Figure \ref{f:Md} plots the number of galaxies classified as a disk, edge-on and with a dust lane as a function of dust mass. Dust masses in these galaxies are typically in a narrow range of $10^{5-6} M_\odot$. Lower amounts of dust are reported for some edge-on and certainly for disk galaxies but dust lanes occur only in a relatively narrow range of dust masses.

\begin{figure}
\includegraphics[width=0.5\textwidth]{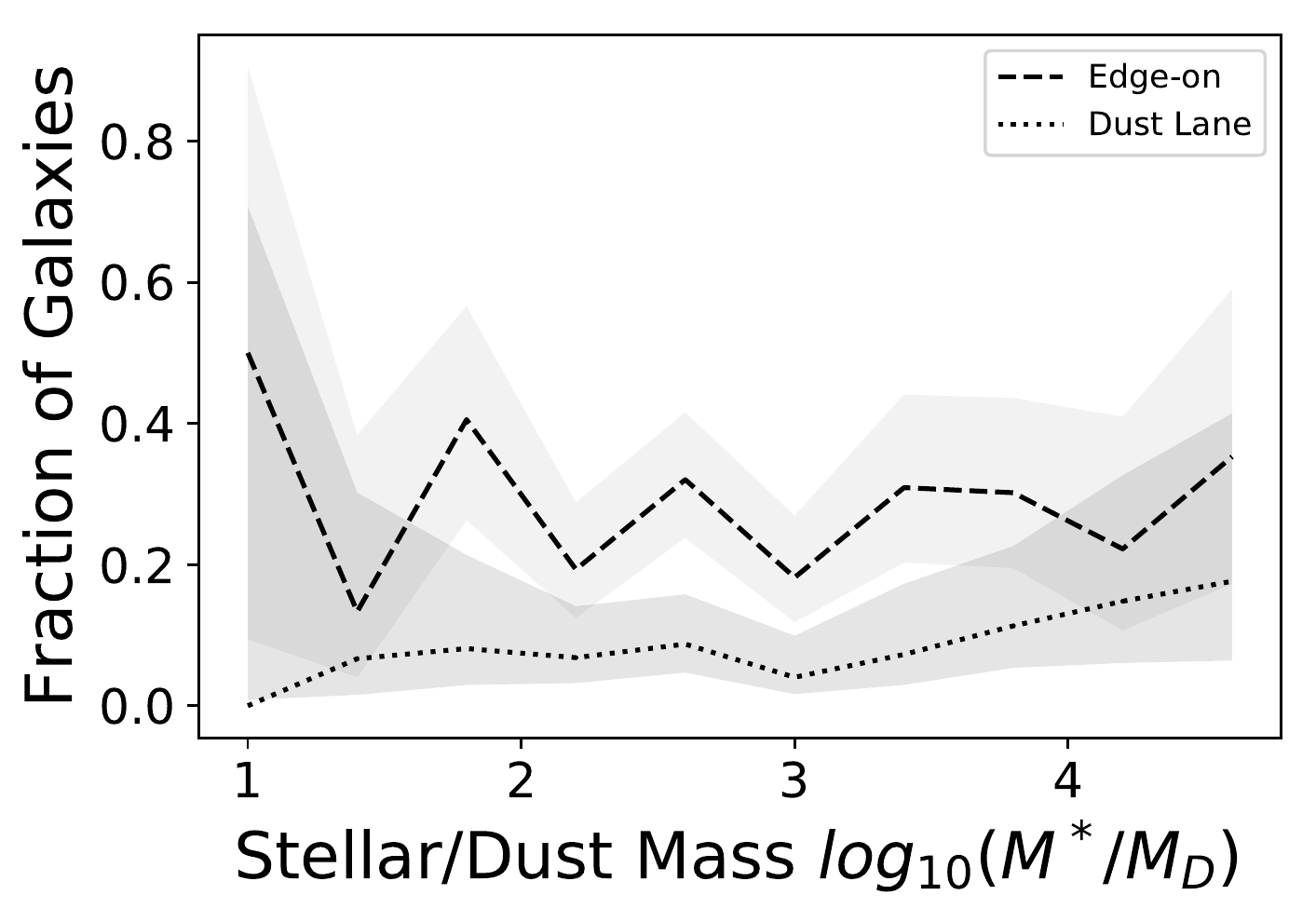}
\caption{\label{f:MstarMd} The fraction of those galaxies identified as disks voted as edge-on (dashed line) and those voted disk, edge-on and showing a dust lane (dotted line) as a function of the ratio of stellar to dust mass ($M^*/M_D$). The fraction of dust lanes rises gently with the stellar to dust mass ratio.}
\end{figure}

\subsubsection{Star/Dust Mass Ratio} \label{s:MstarMd}

A logical follow-up is to explore if the relative masses of stars and dust. One would expect that dust lanes, an inherently indirect measure of dust and reliant on contrast with the surrounding stars to be noticeable, to depend on the ratio of dust to stars.

Figure \ref{f:MstarMd} shows the fraction of edge-on and edge-on with a dust lane as a function of stellar to dust ratio. Dust lanes are identified with only a slightly increased frequency in low (twice as much stellar mass as dust) to high ratio of stellar to dust mass.


\begin{figure}
\includegraphics[width=0.5\textwidth]{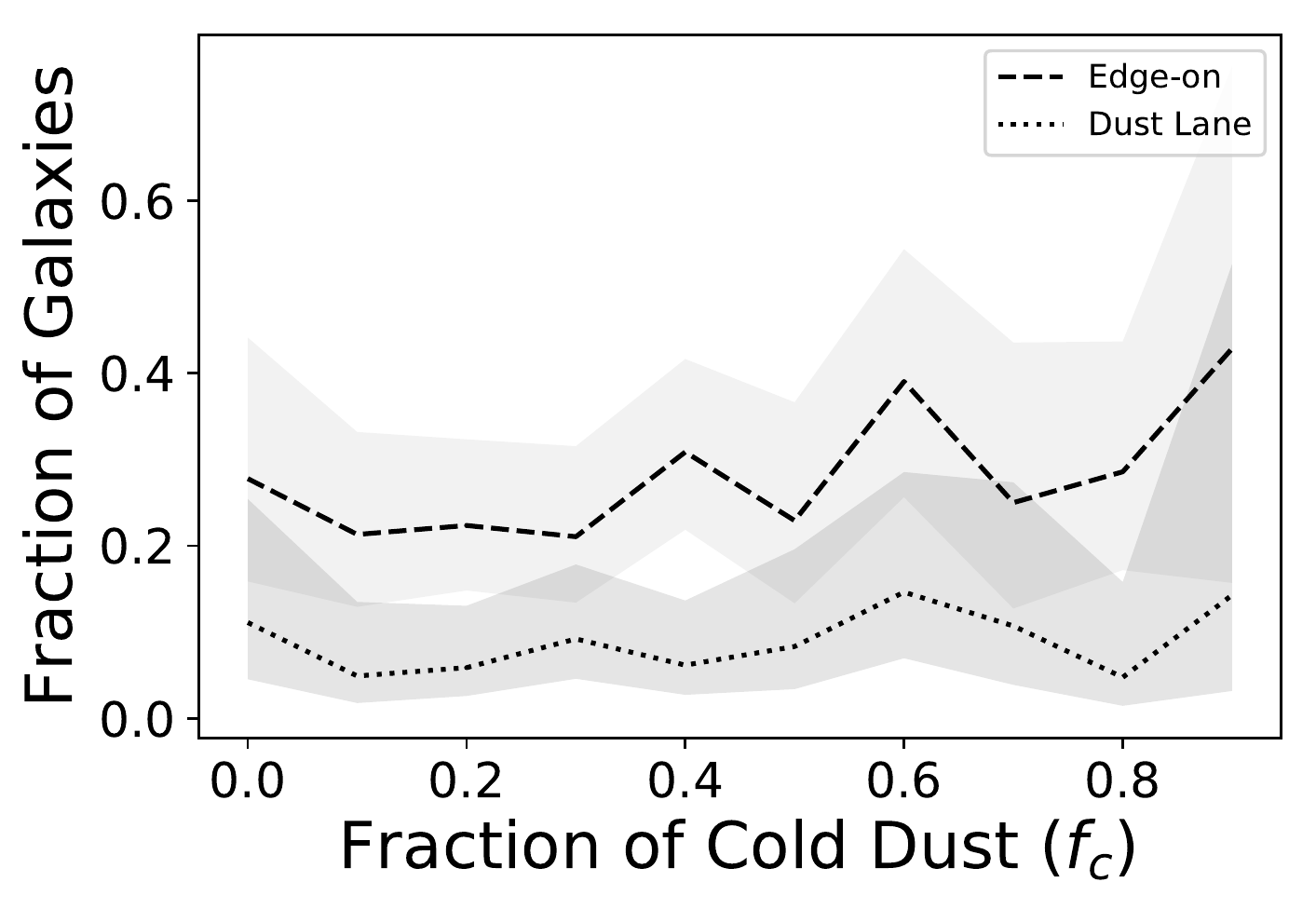}
\caption{\label{f:fc} The fraction of galaxies identified as edge-on (dashed line) and 
with a dust lane (dotted) as a function of cold dust as found by {\sc magphys} by \protect\cite{Driver18}. No relation is visible: dust lane votes remain constant with {\sc magphys} cold dust fraction.}

\end{figure}


\subsubsection{Cold Dust Fraction} \label{s:fc}

Figure \ref{f:fc} shows the histogram of galaxies voted to have a dust lane as a function of {\sc magphys} cold dust fraction ($f_c$). This is the  fractional contribution by cold dust to the dust luminosity of the ambient ISM according to the {\sc magphys} best fit. 
There is no correlation between the fraction of cold ISM identified by {\sc magphys} and the fraction of votes in favor of a dust lane: a cold component can be evident in the SED fit or a dust lane is identified in the images but the two effects do not appear to correlate at all.

\begin{figure}
\includegraphics[width=0.5\textwidth]{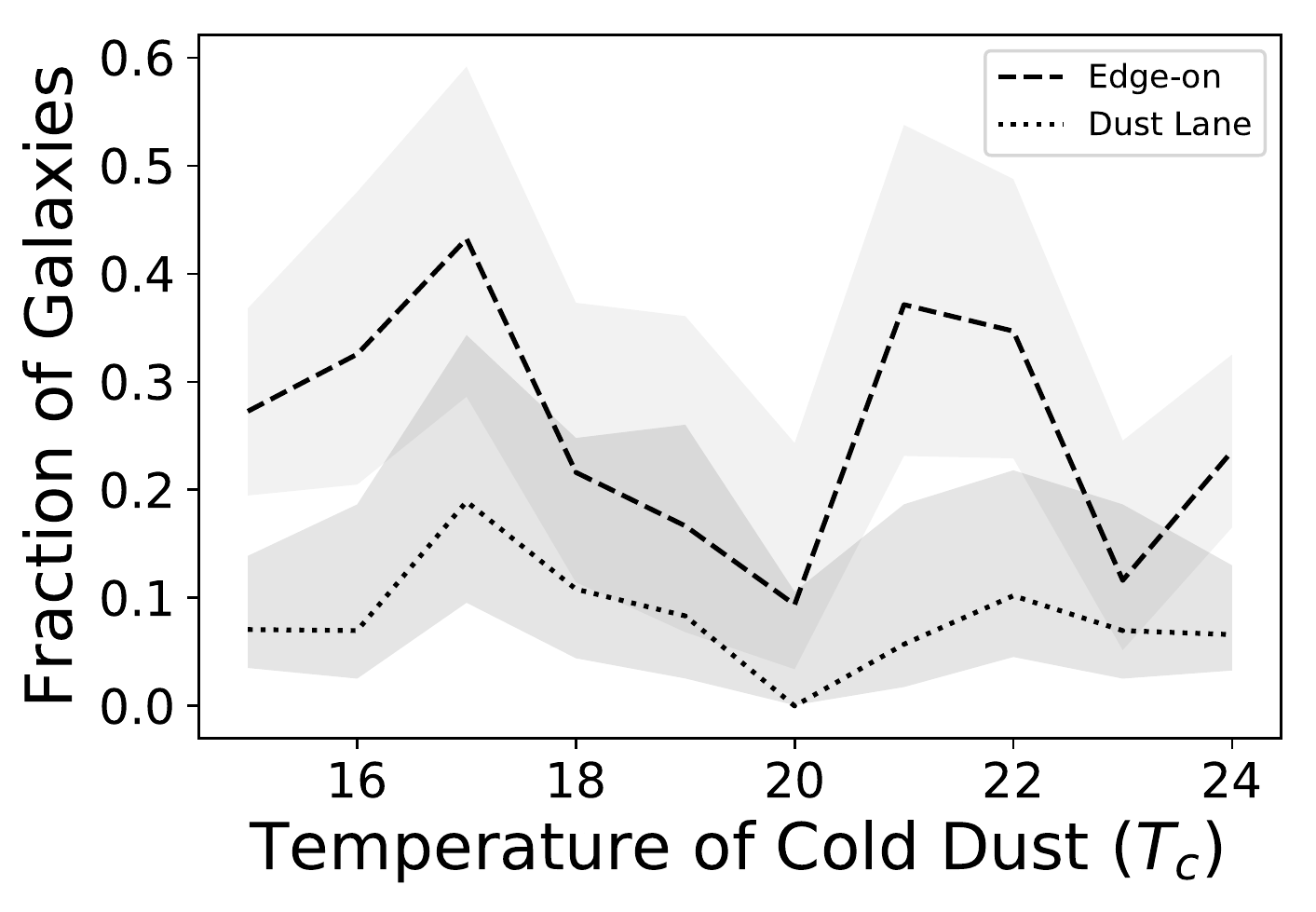}
\caption{\label{f:Tc} The fraction galaxies identified as edge-on (dashed line) and with dust lanes as a function of temperature of cold dust component as found by {\sc magphys} by \protect\cite{Driver18}. No relation is visible; dust lane votes remain constant with cold dust temperature.}
\end{figure}

\subsubsection{Cold Dust Temperature} \label{s:Tc}

Figure \ref{f:Tc} shows the votes for dust lanes as a function of {\sc magphys} cold dust temperature ($T_c$). No dependence on dust temperature is evident. The {\sc magphys} cold dust temperature ($T_c$) refers to the cold component in the diffuse ISM. One would --perhaps naively-- expect there to be a correlation as the colder ISM would sinks to the plane of the disk as dense clumps of ISM, enhancing the dust lane effect and the warmer ISM  is distributed more vertically.

Once a cold dust component is present in the diffuse ISM, it appears decoupled from the identification of a dust lane.

\begin{figure}
\includegraphics[width=0.5\textwidth]{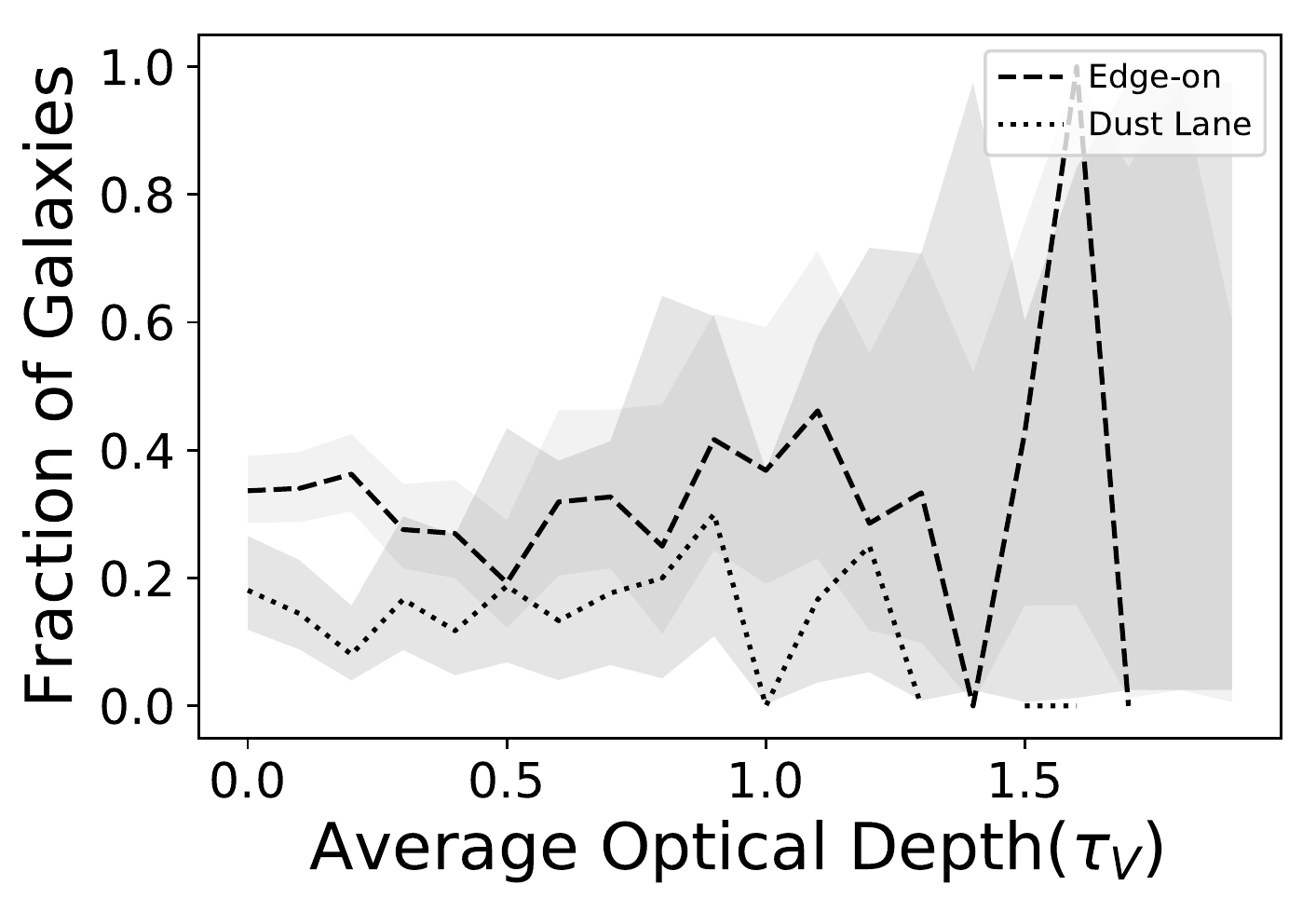}
\caption{\label{f:tau} The average face-on optical depth as determined by {\sc magphys}. The majority of disk galaxies is optically thin according to {\sc magphys}. The fraction of dust lanes declines steadily with the edge-on fraction. }
\end{figure}

\subsubsection{Optical Depth} \label{s:tau}

Figure \ref{f:tau} shows the relation between  face-on optical depth computed by {\sc magphys} and the fraction of dust lane votes. The majority of galaxies in our sample are considered optically thin by {\sc magphys}. Only a few dozen edge-on galaxies are in the optically thick regime ($\tau_V>1$). A dust lane is by definition optically thick and hence this result shows the {\sc magphys} result is mostly based on the light from either side of the dark, optically thick lane. Yet, given that dust lanes can be the integral effect of a diffuse ISM, their presence should be related to the inferred optical depth by {\sc magphys}.
The edge-on galaxies with a dust lane remain a steady fraction of the edge-on voted galaxies as a function of optical depth. Optically thick galaxies present with too low statistics to say much about their dust lane fraction. 

\subsection{Morphology} \label{s:morph}

\cite{Dalcanton04} speculate that the presence of a dust lane is also linked to the oblateness of the disk with flatter disks showing more prevalent dust lanes. In addition, one can expect that bars perturb the molecular cloud arrangement that is responsible for the dark dust lane. 
This can be explored by comparing the dust lanes votes as a function of near-infrared (least influenced by dust) stellar disk appearance, either disk oblateness, ellipticity, or axis ratio (B/A) or the relationship between dust lane votes and votes for different bulge morphologies. 

We selected edge-on disks using the Galaxy Zoo votes with the purpose of comparing them to other morphological features identified and this axis ratio and ellipticity explicitly to test the initial results in \cite{Dalcanton04}. 

\begin{figure*}
\includegraphics[width=0.3\textwidth]{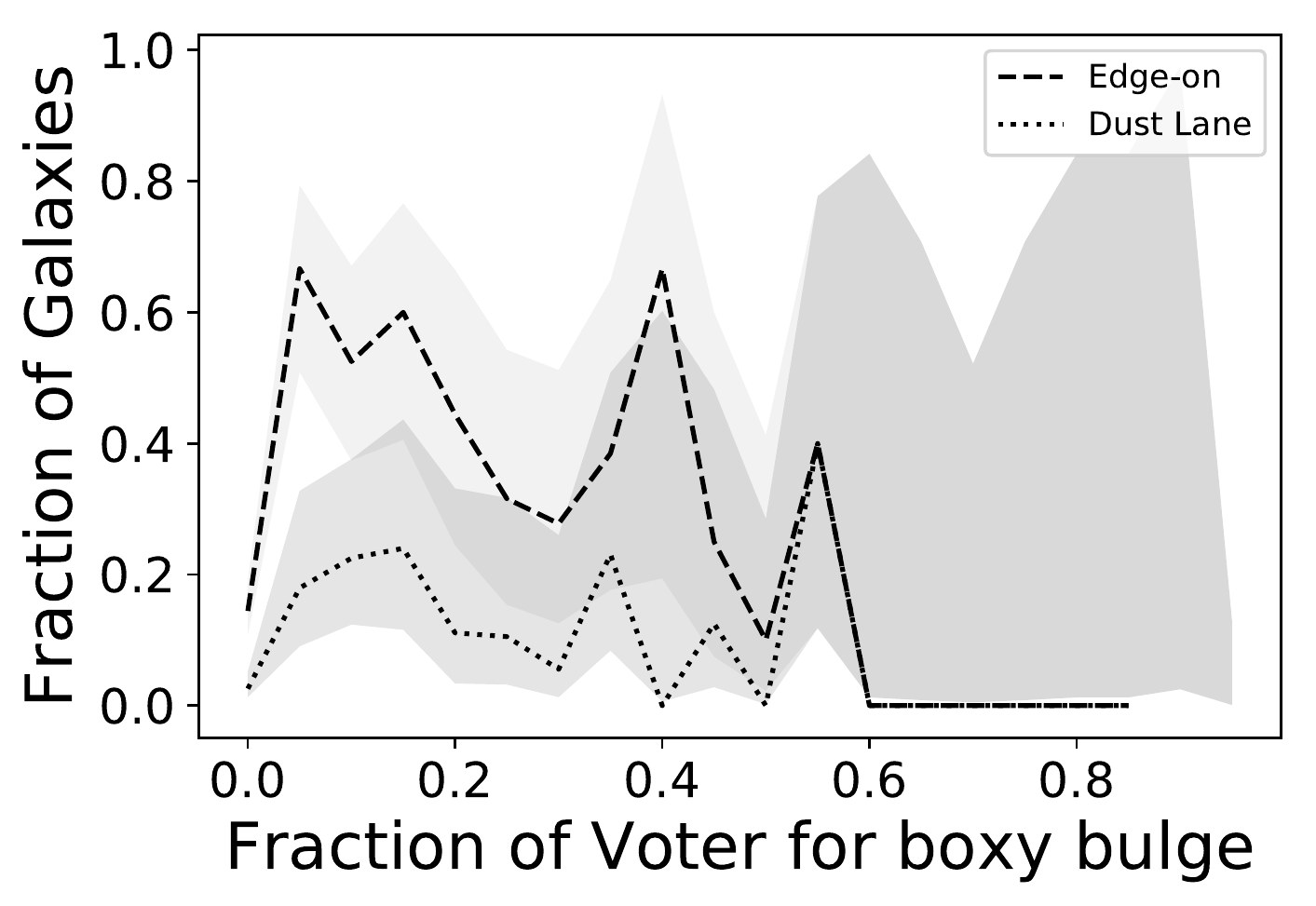}
\includegraphics[width=0.3\textwidth]{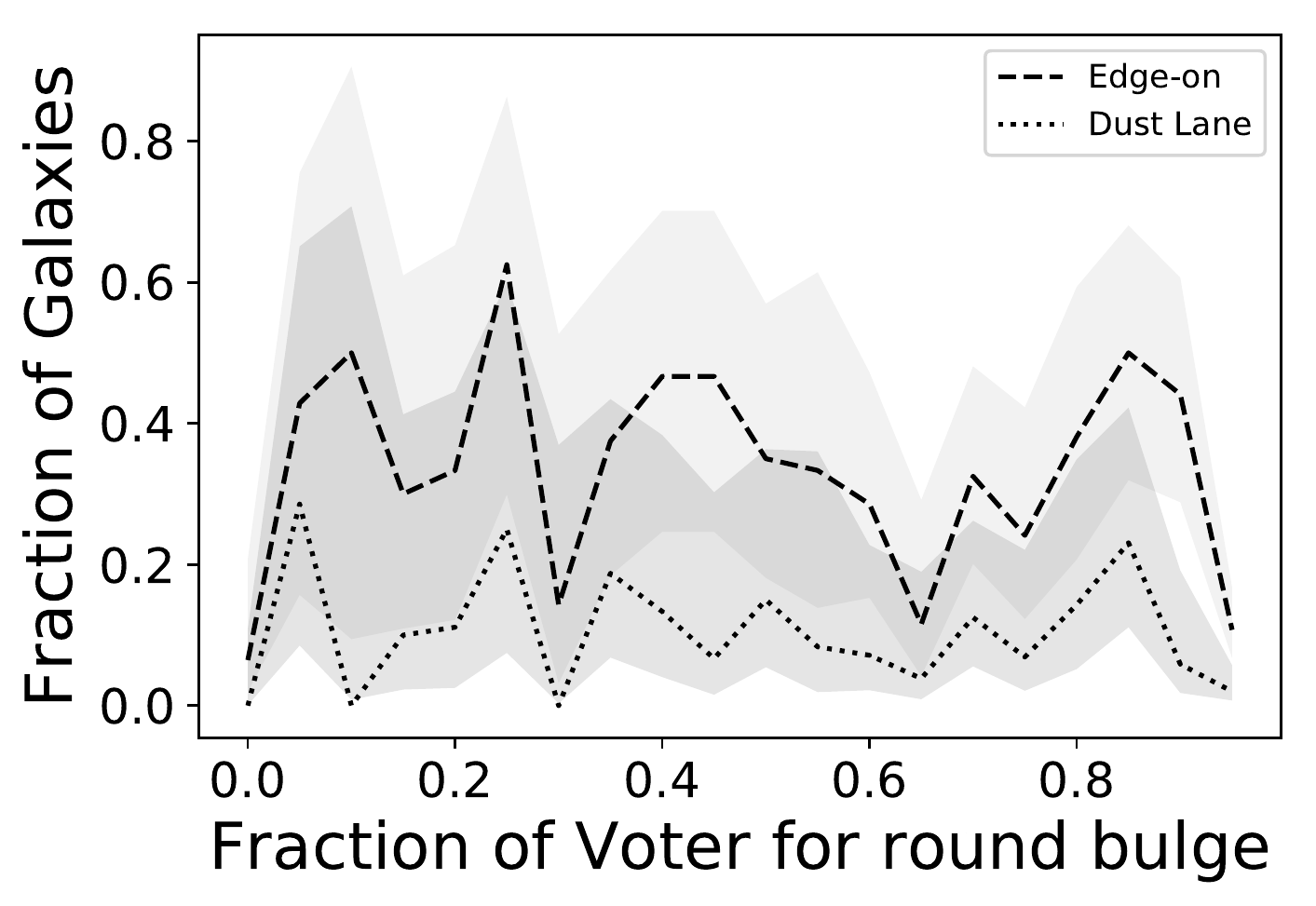}
\includegraphics[width=0.3\textwidth]{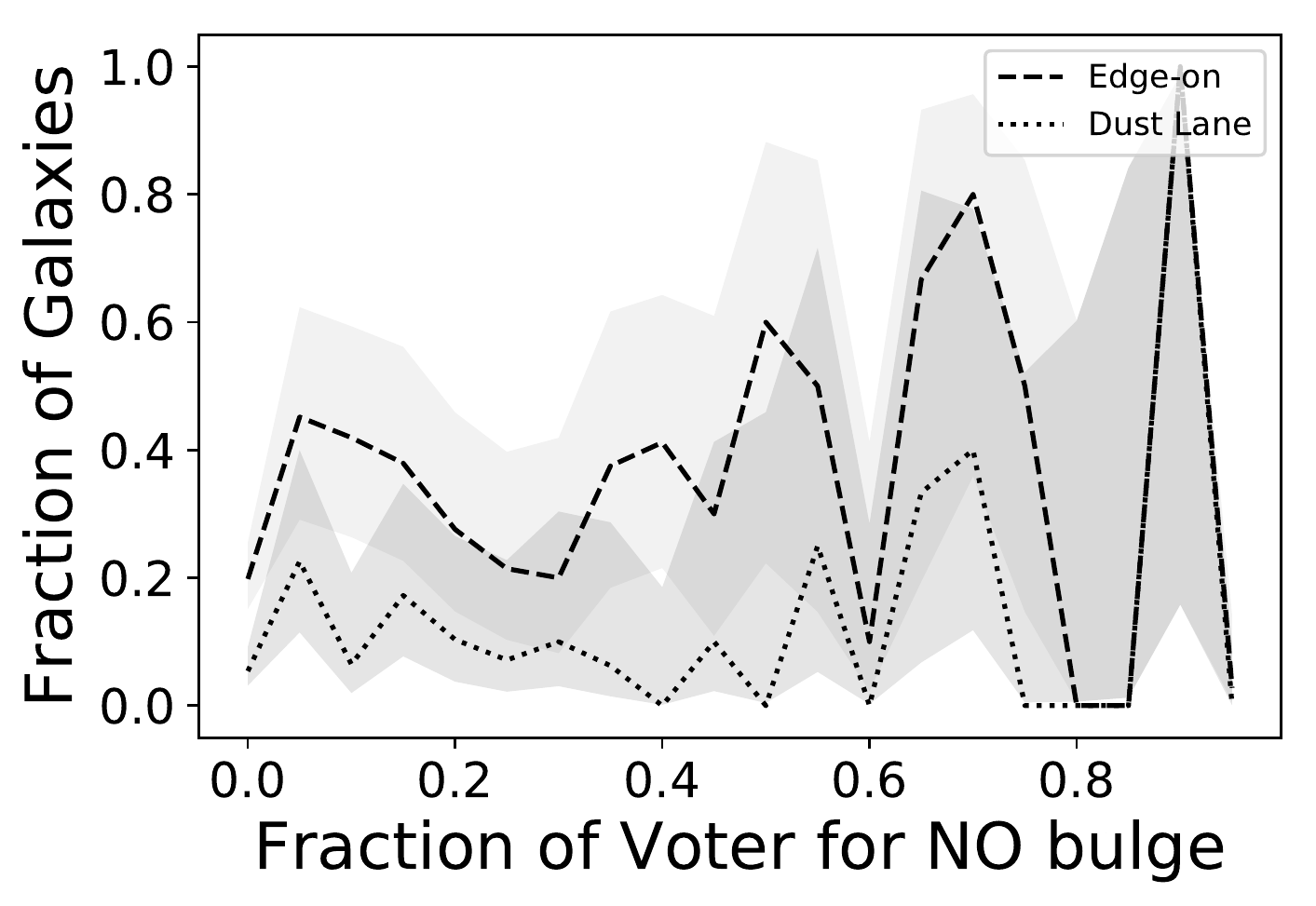}
\caption{\label{f:bulges} The fraction of galaxies voted on as edge-on (dashed), and displaying a dust lane (dotted line) as a function of the fraction of votes in favor of a boxy bulge (left), a round bulge (middle) and no bulge (right).}
\end{figure*}

\subsubsection{Bulges} \label{s:bulges}

Figure \ref{f:bulges} shows the distribution of disk galaxies, galaxies voted as edge-on, and the number of those galaxies with a vote in favor of a dust lane as a function of the fraction of votes in favor of a boxy bulge, a round bulge or no bulge at all. Boxy bulges are seen as evidence for a bar in edge-on galaxies. 

Round bulges at any level of confidence always have the same fraction of edge-on galaxies with a dust lane. The number of galaxies with dust lanes votes anti-correlates with no bulge votes and the number of galaxies with dust lane votes decreases with with increasing voter confidence for a boxy bulge.
Therefore no, or a boxy bulge decreases the chance of a dust lane being identified and a round bulge has no influence on a dust lane identification. This result is a little counter-intuitive as a prominent bulge should aid in highlighting a dust lane bisecting the plane of the disk. If boxy bulges are linked to bars, as they often are in the literature, then this points to a clearing out of dust in the inner disk, resulting in fewer dust lane identifications. The correlation between a lack of dust lanes with a lack of a bulge identification could be a visual selection effect as dust lanes are a little less backlighted in bulgeless galaxies.

\begin{figure}
\includegraphics[width=0.5\textwidth]{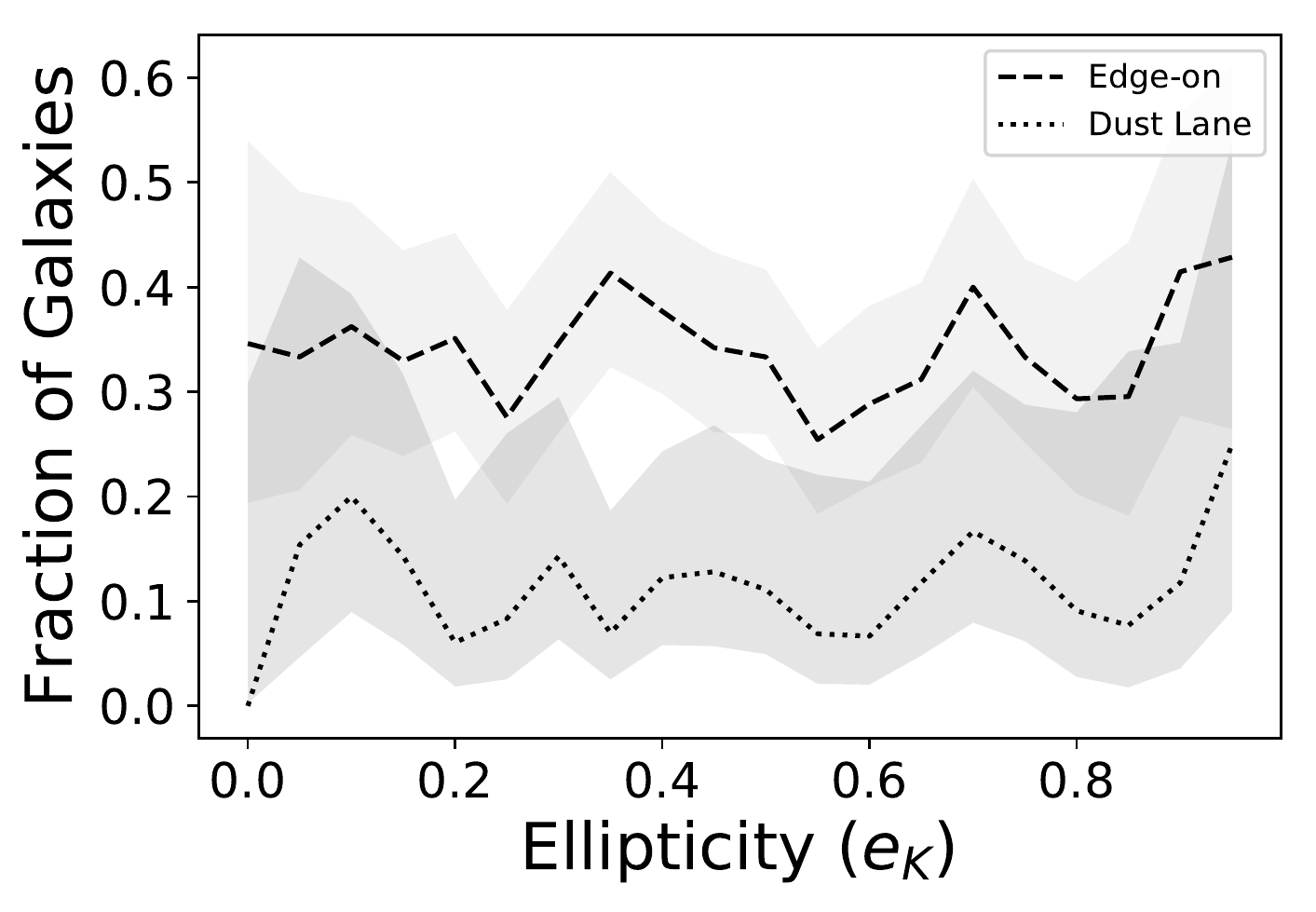}
\caption{\label{f:ell} The fraction of galaxies 
votes as edge-on (dashed), and with a dust lane (dotted) as a function of UKIDDS ellipticity.}
\end{figure}

\subsubsection{Disk Oblateness or Ellipticity} \label{s:Ellipticity}

\cite{Dalcanton04} noted how the edge-on disks display not only dust lanes but a 
flattened stellar disk as well. We use the UKIDSS near-infrared ellipticity to explore 
this observation using the Galaxy Zoo votes for both edge-on disk and dust lanes. 
Figure \ref{f:ell} shows the UKIDSS K-band based ellipticity and the fraction of galaxies 
classified as edge-ons, and the fraction identified with a dust lane. The numbers are remarkably 
steady with ellipticity, showing no preference. Similarly, the Sersic index has little 
influence on the prevalence of dust lanes. 

The lack of dependence on dust lane votes on axis ratio (ellipticity) is surprising given 
the strong rationale \cite{Dalcanton04} make but we should bear in mind that both ellipticity and
dust lane identification are strongly dependent on the inclination of the disk. Only near 
perfectly edge-on will both the dust lane be unequivocal and the ellipticity most extreme. 
Because we use the Galaxy Zoo edge-on identification, this result may have been diluted 
by not perfectly edge-on systems with more median ellipticity values and more difficult to 
identify dust lanes.

A second limitation is that the optical KiDS data analyzed by the Galaxy Zoo is both deeper and higher spatial resolution than the UKIDS K-band data, which likely results in rounder (lower ellipticity) K-band measurements for these galaxies. 
The combined effects have likely smoothed out any dependence. 

\begin{figure}
\includegraphics[width=0.5\textwidth]{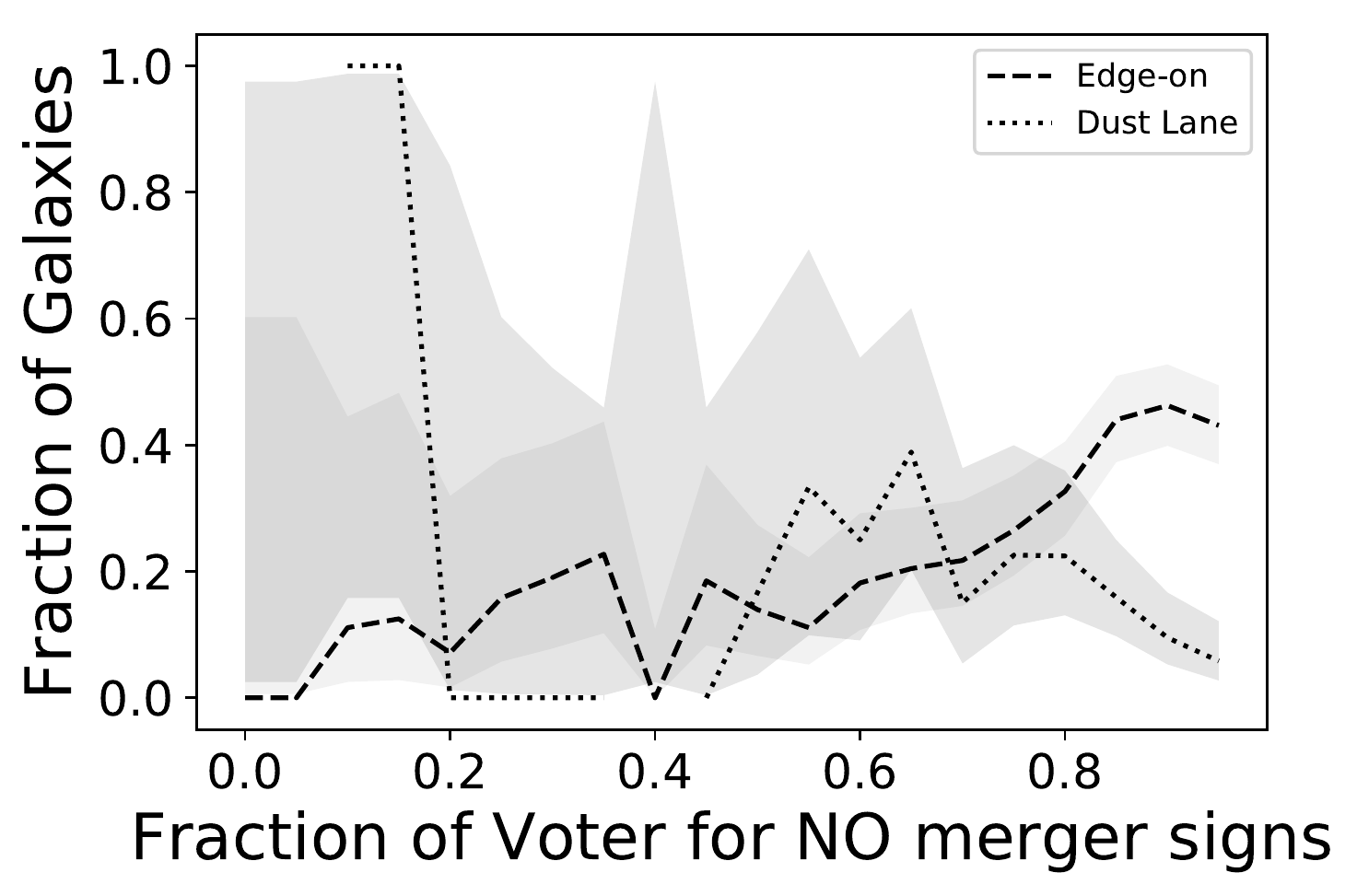}
\caption{\label{f:env} The fraction of galaxies voted edge on (dashed line) and with a dust lane (dotted line) as a function of the fraction of votes for showing no merger signs. The fractions strongly suggests that dust lanes are preferentially identified in galaxies where there no signs of a merger are found. }
\end{figure}

\subsection{Environment}

The question whether there are signs of an ongoing interaction or merger and the presence of a dust lane are now separated (questions Figure \ref{f:tree}). This opens the possibility to explore whether tidal effects of a merger influence the presence of a dust lane. \cite{Holwerda13a} find that in UGC 3995, a mid-stage interaction, the diffuse component of dust has been mostly destroyed or swept up into dense structures. This provides a hint that the dusty ISM is radically rearranged in the early stages of a merger or interaction. 

The fraction of votes in favor of a dust lane and no interaction appear to be correlated (Figure \ref{f:env}). Either an interaction distracts from the identification of dust lanes, or dust lanes are perturbed/removed by an interaction or a combination of these effects.

\section{Discussion} \label{s:disk}

Dust lanes are common in edge-on galaxies, so much so that votes for their presence need not be numerous, they are often considered unremarkable by classifiers. The presence of dust lanes depends most strongly on stellar mass. Dust lanes are increasingly identified where the {\em relative} dust mass is smaller compared to the stellar disk. 

Dust lane presence depends on the type of bulge visible in the edge-on disk, with votes for boxy bulges --thought connected to the presence of a bar-- anti-correlated with the presence of dust lanes. Similarly, the presence of dust lanes anti-correlates with signs of interaction, recent or ongoing. 

Our results point towards a scenario where dust lanes need unperturbed disks of a certain mass to show clearly. This is in line with the discussion by \cite{Dalcanton04}, where the vertical stability of the disk is such that dusty ISM clouds sink to the central plane as well as the result by \cite{Holwerda13a} that shows the diffuse dust is removed/swept up by an interaction in the early stages.

The lack of a correlation with the {\sc magphys} parameters and the occurrence of dust lanes in the Galaxy Zoo classifications is puzzling. 
We note that the {\sc magphys} values for individual galaxies still hold large uncertainties in the derived parameters \citep{Wright18}.
The thin, cold, dusty ISM responsible for the dust lane should logically be associated with one of the two components used in {\sc magphys}; the cold, compact one, not the warm component heated by star-formation. Most vexingly, it correlates with neither clearly.

One can look to the radiative transfer results to interpret these result: the optically thick component in the plane of the disk is neither warm nor cold, nor is it exclusively in dense clouds or the diffuse component best probed with {\sc magphys}. 

The cold dusty clumps may occur in the plane of the stellar disk, they are deeply embedded in this disk: their contribution to the line-of-sight optical depth in the edge-on perspective is smoothed out by the nearby parts of the stellar disk. Equally, the warmer diffuse component is not optically thick but it's effect is felt over a larger path along the line of sight, resulting in an equal contribution to the dust lane. 

The stellar-to-dust mass ratio effect is similarly counter-intuitive but dust lanes need a stellar disk to contrast against. 


Following this, we note that a dust lane requires a minimum mass of dust $M_D \sim 10^5 M_\odot$ but it can occur in any stellar mass disk (above $10^9 M_\odot$) and any oblateness.

\section{Conclusions} \label{s:concl}

Using the Galaxy Zoo classifications of the KiDS data overlapping with the GAMA equatorial fields, we examine the frequency of dust lanes in edge-on galaxies and relate them to other observables of the galaxies. We find the following:

\begin{itemize}
\item Dust lanes are seen to occur most frequently in above a stellar mass of $10^{8.5} M_\odot$. This corresponds reasonably to the one found by \cite{Dalcanton04} for stellar mass: $10^{9.8} M_\odot$ corresponding to $v_{rot} \sim 120 $ km/s), (Figure \ref{f:hist:mstar})
\item The occurrence of a dust lane appears poorly or not at all correlated with any {\sc magphys} dust parameters, (Figures  \ref{f:fc}, \ref{f:Tc}, \ref{f:tau}), indicating the dust lane is not associated with either dust component alone but a cumulative effect of all the dust in the disk.
\item The dust lanes occur in galaxies with a minimum dust mass of $M_D \sim 10^5 M_\odot$ (Figure \ref{f:Md}) but show a wide range of stellar to dust mass ratios (Figure \ref{f:MstarMd}). Dust lanes may be identified more prevalently in relatively more massive stellar disks.
\item The identification of a boxy bulge and the presence of a dust lane appears anti-correlated (Figure \ref{f:bulges}), suggesting boxy bulges (bars) are involved in sweeping clear their inner disk of dust.
\item Dust lanes and signs of interaction anti-correlate (Figure \ref{f:env}), confirming a scenario where the dust ISM is rearranged early in a galaxy-galaxy interaction.
\end{itemize}

Future work on the frequency of dust lanes in edge-on galaxies can employ the full analysis of the Galaxy Zoo classifications of the Dark Energy Sky Survey images or follow-up Galaxy Zoo projects to answer questions on the size and morphology of the dust lanes. Key to discriminating whether this is a sharp transition at $10^9 M_\odot$ stellar mass or a smooth one will be much improved statistics on dust lane frequency in lower mass disk galaxies. The combination of the higher resolution and statistics make that practical with WFIRST or perhaps LSST.

\section*{Acknowledgements}

This publication has been made possible by the
participation of more than 20000 volunteers in the Galaxy Zoo project. Their contributions are individually acknowledged at \url{http://authors.Galaxy Zoo.org/}.

This research has made use of the NASA/IPAC Extragalactic Database (NED) which is operated by the Jet Propulsion Laboratory, California Institute of Technology, under contract with the National Aeronautics and Space Administration. 
%
%
This research has made use of NASA's Astrophysics Data System.
This research made use of Astropy, a community-developed core Python package for Astronomy \citep{Astropy-Collaboration13a}. This research made use of matplotlib, a Python library for publication quality graphics \citep{Hunter07}. PyRAF is a product of the Space Telescope Science Institute, which is operated by AURA for NASA. This research made use of SciPy \citep{scipy}.

%

\end{document}